\documentclass[a4paper,fleqn,usenatbib,useAMS]{mnras}
\usepackage{graphicx}	
\usepackage{amsmath}	
\usepackage{amssymb}	

\title[Impact of cosmological cascades]{
   Physics of cosmological cascades and observable properties
}

\author[T. Fitoussi et al.]{
   T. Fitoussi$^{1,2}$\thanks{E-mail: thomas.fitoussi@irap.omp.eu},
   R. Belmont$^{1,2}$\thanks{E-mail: renaud.belmont@irap.omp.eu},
   J. Malzac$^{1,2}$,
   A. Marcowith$^{3}$,
   J. Cohen-Tanugi$^{3}$,
  \newauthor and P. Jean$^{1,2}$ \\
   $^{1}$ Universit\'e de Toulouse; UPS-OMP; IRAP; Toulouse, France \\
   $^{2}$ CNRS; IRAP; 9 Av. colonel Roche, BP 44346, F-31028 Toulouse cedex 4, France \\
   $^{3}$ Laboratoire  Univers  et  Particules  de  Montpellier,  Universit\'e de Montpellier,
   CNRS/IN2P3, Montpellier, France\\
}

\date{Accepted XXX. Received XXX.; in original form \today}
\pubyear{\today}

\begin{document}
\label{firstpage}
\pagerange{\pageref{firstpage}--\pageref{lastpage}}
\maketitle

\begin{abstract}
   TeV photons from extragalactic sources are absorbed in the intergalactic medium and
   initiate electromagnetic cascades. These cascades offer a unique tool to probe the
   properties of the universe at cosmological scales. We present a new Monte Carlo
   code dedicated to the physics of such cascades. This code has been tested against both
   published results and analytical approximations, and is made publicly available. Using
   this numerical tool, we investigate the main cascade properties (spectrum, halo
   extension, time delays), and study in detail their dependence on the physical
   parameters (extra-galactic magnetic field, extra-galactic background light, source 
   redshift, source spectrum and beaming emission). The limitations of analytical 
   solutions are emphasised. In particular, analytical approximations account only for 
   the first generation of photons and higher branches of the cascade tree are neglected. 
\end{abstract}


\section{Introduction}

The Universe is opaque to gamma-rays. Very high-energy photons from extragalactic sources 
are absorbed by the ambient soft radiation and converted into electron-positron pairs
\citep{gould_opacity_1967,wdowczyk_primary_1972}. These leptons are deflected by the
ExtraGalactic Magnetic Field (EGMF) and cool through inverse Compton scattering, 
producing new gamma-rays that may, in turn, be absorbed. The observable properties of the 
resulting electromagnetic cascade depend on the characteristics of the intergalactic 
medium. The development of cascades has three main observable 
effects. First, the source spectrum is altered because each high energy TeV photon is 
reprocessed into thousands of GeV photons \citep{protheroe_effect_1986,
roscherr_constraining_1998,aharonian_origin_2002,neronov_evidence_2010}. Second, due to 
the deflection of leptons by the EGMF, new gamma-rays are emitted along different lines of 
sight, so that a point source may appear as extended \citep{aharonian_very_1994,
eungwanichayapant_very_2009}. Third, as leptons are deflected, cascade photons travel a 
longer distance and arrive with a significant time delay, as compared to unabsorbed, 
primary photons \citep{kronberg_time_1995,plaga_detecting_1995,
ichiki_probing_2008,murase_probing_2008,takahashi_detectability_2008}. 

Observation (or non-detection) of electromagnetic cascades is crucial to several 
astrophysical issues. It offers a unique tool to probe the intergalactic medium, 
especially the Extragalactic Background Light (EBL) and the EGMF. The background photons 
involved in the cascades have two distinct origins. Inverse Compton scattering mainly 
occurs on photons from the Cosmic Microwave Background (CMB) while high energy gamma-rays 
are mostly absorbed by the EBL of stars and dust, which 
extends from infrared to ultraviolet. Our knowledge of the EBL is limited. Direct 
observations at these wavelengths are very inaccurate due to strong contamination from 
the zodiacal light. The predictions of the different models proposed in the 
literature can differ by up to an order of magnitude, depending on wavelength and 
redshift \citep[see Fig.~\ref{fig:EBL_CMB}]{franceschini_extragalactic_2008,
dominguez_extragalactic_2011,finke_modeling_2010,kneiske_lower-limit_2010,
gilmore_semi-analytic_2012}. Absorption of the gamma ray spectrum of high-energy sources 
provides unequal constrains on the EBL \citep{stecker_tev_1992}. 

Recently, cosmological cascades were also used to probe the properties of the extragalactic
magnetic field, the origin of which is still debated \citep{durrer_cosmological_2013}. A 
primordial magnetic field could have been generated during inflation or during the phase 
transition when electroweak and QCD forces decoupled. This field would have remained 
unaffected during the evolution of the extragalactic medium. Alternatively magnetic 
fields generated by galaxies during large scale structure formation could have propagated 
in the intergalactic medium through plasma jets. Depending on the properties of the field 
generation and evolution, its value $B$ is expected to lie in the range $10^{-17}$ to 
$10^{-9}$ Gauss \citep{essey_determination_2011,finke_constraints_2015},
with coherence length $\lambda_B$ (scale of de-correlation of two nearby 
field lines) between $10^{-6}$ and $10^4$ Mpc. Electromagnetic cascades represent a 
unique tool to probe the intergalactic magnetic field \citep{aharonian_ultrahigh_1985} 
when conventional methods such as Faraday rotation cannot be applied. 
\cite{neronov_method_2007} and \cite{elyiv_gamma-ray_2009} have suggested to measure the 
extension of pair halos to probe the EGMF. Indeed 
\cite{neronov_degree-scale_2010} demonstrated that for a strong enough magnetic field, 
halos in the GeV energy band can remain long after the TeV blazar's end of activity. 
Alternatively, the spectral analysis can also provide constraints on the EGMF 
\cite[e.g.][]{davezac_cascading_2007,neronov_evidence_2010, kachelries_constraints_2010}.
Although most studies focus on the average intensity and coherence length of the 
field, it has been shown recently that anisotropies in the images of pair halos could 
also provide crucial information on the magnetic helicity \citep{long_morphology_2015,
batista_probing_2016}.

In the past years, all three effects have been searched intensely in the data of the 
space gamma-ray telescope {\it Fermi} and of Cherenkov air telescopes such as MAGIC, 
H.E.S.S, or VERITAS. However none of the methods has provided undisputed evidence yet. 
Cascade contribution to the GeV spectrum has mostly provided upper limits, and most 
Blazar observations remain compatible with no cascade emission 
\citep{arlen_intergalactic_2014}. Such constraints however provide lower limits on the 
EGMF intensity \citep{neronov_evidence_2010}. No time delay has been clearly detected 
either, which also provides lower limits on the amplitude of the random component of the 
magnetic field \citep{neronov_no_2011}. Detection of pair halos requires a very accurate 
modelling of the instrument point spread function (PSF) and has not given undisputed 
results either \citep{krawczynski_x-ray/tev-gamma-ray_2000, aharonian_search_2001,
abramowski_search_2014, prokhorov_search_2016}. A detection in Fermi-LAT data sets was 
claimed recently \citep{chen_search_2015}, but has not been confirmed by the {\it Fermi} 
collaboration yet. Much better constraints are expected from CTA 
\citep{meyer_sensitivity_2016}. 

Regardless of the detection method, a deep understanding of the cascade physics is 
crucial to interpret observational data. In the past decade, the cascade physics has been 
investigated through fast, analytical (or semi-analytical) methods that allow to quickly 
cover a large parameter space, and Monte Carlo simulations. Although much slower, the 
latter have proven to be mandatory to derive quantitative results and to interpret 
precise observations. Several codes have been developed over the years but only the most 
recent include the cosmological expansion on the particle trajectory
\citep{taylor_extragalactic_2011,kachelriess_elmag_2012, arlen_intergalactic_2014,
settimo_propagation_2015}. To our knowledge, only one is publicly available 
\citep[ELMAG][]{kachelriess_elmag_2012} but the lepton trajectories in the magnetised, 
intergalactic medium is treated in a simple, 1D, diffusion approach. 
In this paper we present a new Monte Carlo code that is publicly 
available\footnote{\url{https://gitlab.com/tfitoussi/cascade-simulation}}. 
This code is dedicated to cascades induced by high-energy photons (or leptons) and does not take 
into account hadronic processes \citep[see e.g. ][for results on cosmic-ray induced 
cascades]{oikonomou_synchrotron_2014,essey_determination_2011}.
It computes the physics of leptonic cascades at the highest level of precision and with 
the fewest approximations. Using this code, we present a systematic exploration of the 
parameter space.

In Section~\ref{sec:cascades} we present the basic analytical theory of cosmological 
cascades and simple analytical estimates of their observables. The results of our 
code are presented in Section~\ref{sec:simple_case} and are tested against analytical 
approximations and other published numerical results. The last sections of the paper 
is devoted to an exploration of the parameter space. We study the impact of the source 
properties (redshift, spectrum, anisotropy) in Section \ref{sec:source_profile}. Then, in 
Section~\ref{sec:intergalactic_medium}, we explore the effects of the intergalactic 
medium (EBL, EGMF). Technical aspects of the code are presented in Appendix
\ref{appendix:code}.

\section{Physics of cosmological cascades}
\label{sec:cascades}

Cosmological electromagnetic cascades involves three main processes: pair 
production through photon-photon annihilation, inverse Compton scattering, and 
propagation of charged particles in a magnetised, expanding universe. All other processes 
are negligible. In  particular, as long as primary photons do not exceed 100 TeV and 
the extragalactic magnetic field remains below $B=10^{-10}$ G, synchrotron cooling is 
orders of magnitude weaker than Compton cooling, and synchrotron photons only contribute 
at low energy, below the infrared range ($< 0.02$ eV).
This section presents a simple analytical view of the cascade physics.

\subsection{Propagation of particles in a magnetised, expanding Universe}

Cosmological cascades develop on kpc to Gpc scales. On the largest scales, the geometry 
and the expansion of the universe must be taken into account. Throughout this paper we 
assume a $\Lambda$-CDM model. A complete description of particle trajectories can be 
found in appendix \ref{appendix:travel}. However a few important points must be noted 
here. First, in an expanding Universe and in absence of any interaction, the particle 
(photons and leptons) momentum $p$ evolves with redshift $z$ as $p \propto (1+z)$, also 
meaning that their energy continuously decreases with time. In the specific case of 
photons, the energy scales with momentum: $E_\gamma \propto p \propto (1+z)$, providing the 
well-known cosmological redshift. The cosmological evolution of lepton energy is slightly
more complex. However, in the limit of highly relativistic particles, it also scales as 
$(1+z)$. 

The propagation of leptons is also affected by the extragalactic magnetic field 
(EGMF). In this work, we assume that no field is created or dissipated in the 
cosmological voids, and that it is simply diluted as the universe expands: 
$B(z)\propto (1+z)^2$ \citep[see][eq. 22]{durrer_cosmological_2013}. In that case, 
the Larmor radius evolves as:
\begin{equation} \label{eqn:RL}
   R_L = \frac{E_e}{e_cB (1+z)}
      \approx 1.1  ~(1+z)^{-1} \left( \frac{E_e}{1\rm{TeV}} \right)
      \left( \frac{B}{10^{-15}\rm{G}} \right)^{-1}  \textrm{Mpc},
\end{equation}
where $e_c$ is the lepton charge, $E_{e}$ and $B$ are respectively the lepton energy and 
the EGMF intensity at $z=0$. In comoving coordinates, this means that  the comoving 
Larmor radius $ R_L (1+z)$ is constant and that for a uniform magnetic field, the 
perpendicular comoving motion of leptons is a pure circle. When Compton losses are 
included, lepton trajectories become converging spirals.

In practice, the cosmological magnetic field is expected to be highly turbulent 
\citep{caprini_gamma-ray_2015,durrer_cosmological_2013}. Although it should be described 
by a full turbulent spectrum, its properties are often characterised by its intensity $B$ 
and its coherence length $\lambda_B$. In this paper, we will consider that the field 
structure can be modelled by uniform magnetic cells of same intensity and size 
$\lambda_B$ but with random orientations. Inside a particular cell, the lepton
comoving trajectories become simple helicoidal trajectories. 

\subsection{Photon absorption by the EBL}

\begin{figure} \centering
   \includegraphics[width=\columnwidth]{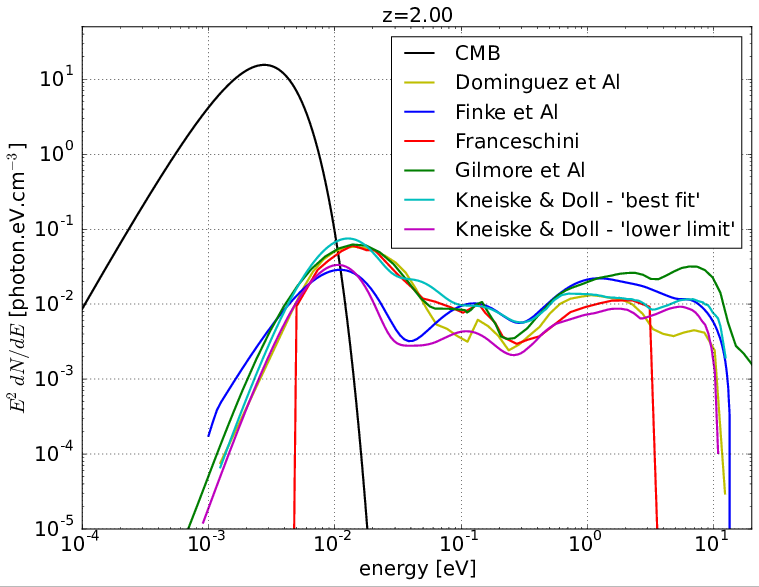}
   \caption{Comoving spectral energy distribution of target photons at a redshift $z=2$, 
   including the Cosmological Microwave Background (black line) and different models 
   of EBL (color lines).}
   \label{fig:EBL_CMB}
\end{figure}

High-energy photons (of energy $E_\gamma$) annihilate with soft, ambient photons. The 
annihilation cross section being maximal close to the threshold, the interaction is 
most efficient with soft photons of energy $\sim (m_ec^2)^2/E_{\gamma}$ where $m_e$ is 
the lepton mass and $c$ is the speed of light. This explains why TeV photons are 
absorbed preferentially by eV photons of the EBL \citep{gould_opacity_1967}. 
Fig.~\ref{fig:EBL_CMB} shows six models of EBL that can be found in the literature 
\citep{franceschini_extragalactic_2008,dominguez_extragalactic_2011,finke_modeling_2010,
kneiske_lower-limit_2010,gilmore_semi-analytic_2012} and illustrates the uncertainty 
of EBL intensity and spectrum. It can be seen that at $z=2$, the EBL photon densities 
can differ by one order of magnitude from one model to the other.

\begin{figure} \centering
    \includegraphics[width=\columnwidth]{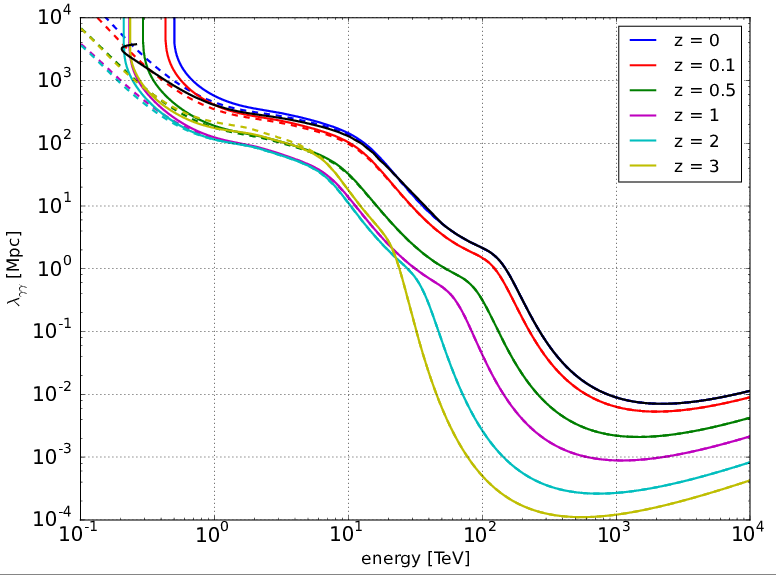}
   \caption{Gamma-ray annihilation mean free path $\lambda_{\gamma\gamma} = c 
      t_{\gamma\gamma}$ (where $t_{\gamma\gamma}$ is the mean cosmic time between two 
      interactions) as a function of their initial energy, and for different emission 
      redshifts (solid lines). For comparison, the mean free path for a static universe 
      (with properties frozen at their values at the initial redshift) is shown in dashed 
      lines. The black line shows the thin/thick transition where the mean free path 
      equals the source distance $D_s$ and where the energy equals the absorption energy 
      $E_{\rm abs}$.}
   \label{fig:lambda_gg}
\end{figure}

The photon mean annihilation distance is plotted in Fig.~\ref{fig:lambda_gg} as a 
function of the initial photon energy, assuming the EBL model from 
\cite{dominguez_extragalactic_2011}. The solid lines show the results for an 
expanding universe while the dashed lines show the results for a static universe. 
Below 1 TeV, the absorption mean free path quickly becomes larger than the typical 
distance of targeted sources (100 Mpc to Gpc), so that only TeV photons are significantly 
absorbed. Below 200-300 GeV, photons tend to travel over such large distances that two 
cosmological effects work in concert to produce a diverging mean free path. First, the 
target photon density vanishes as $(1+z)^3$ as the universe expands (horizon event). 
Second, gamma-rays photons are more and more redshifted before reaching the next 
annihilation point, requiring higher and higher energy target photons. As the EBL photon 
density drops at 10 eV, the mean free path diverges at low energy. Photons emitted at low 
redshift, with an energy of 1~TeV travel a few hundred Mpc before producing pairs. This 
distance decreases quickly at higher energy to reach few Mpc to few kpc. Considering 
a blazar like Mrk421 ($z=0.0308$, 135~Mpc) emitting very high-energy photons up to 
100~TeV, primary gamma-rays are typically absorbed over a distance of a few Mpc.

\begin{figure} \centering
   \includegraphics[width=\columnwidth]{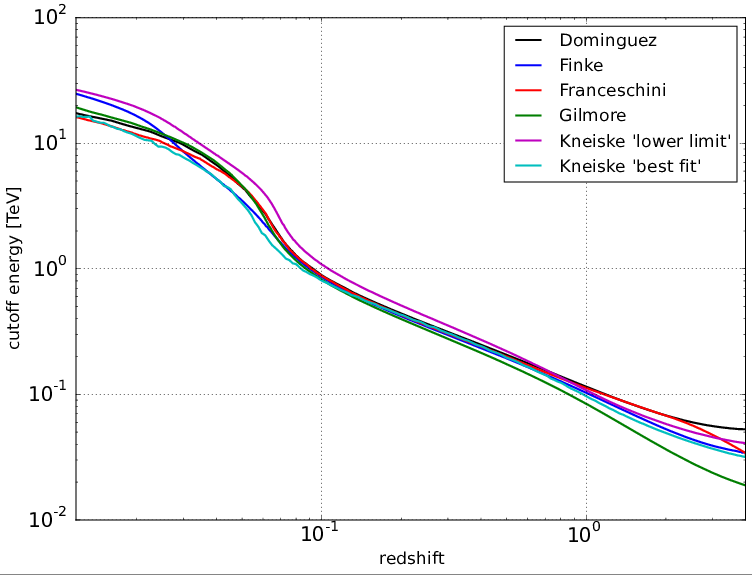}
   \caption{Cut-off energy $E_{\rm cut}$ observed at $z=0$ as a function of the source 
   redshift for different EBL models. }
   \label{fig:cutoff_energy}
\end{figure}

Since the density of the EBL photons decreases with their energy, gamma-ray absorption 
is more efficient at high energy. The transition from optically thin to optically thick 
photon-photon absorption where the radiation becomes fully absorbed occurs at an initial 
energy $E_{\rm abs}=(1+z)E_{\rm cut}$ where $E_{\rm cut}$ is the corresponding energy 
cut-off observed in the spectra at $z=0$. Fig.~\ref{fig:cutoff_energy} shows the cut-off 
energy as a function of source redshift. Distant sources are more absorbed and their 
absorption occurs at lower energy. Significant differences are observed between the 
different EBL models. At large redshift ($z>1$), the different EBL models are very 
different, and differences up to a factor 6 are observed in the cutoff energy. At lower 
redshifts the EBL models are consistent with each other and only differ by a factor 2 at 
the lowest energies as, at such low distance, gamma-gamma absorption occurs mainly above 
10 TeV. The effects of the EBL model on the cascades will be discussed in section 
\ref{sec:EBL}.

\subsection{Compton scattering by the CMB}

High-energy leptons Compton up-scatter soft ambient photons to gamma-ray energies. In 
the Thomson regime, the Compton cross-section does not depend on the energy, and the 
scattering rate scales linearly with the target photon number density, hence Compton 
scattering mostly occurs on CMB photons. The CMB is modelled by a blackbody with a
temperature $T_{\rm cmb} = (1+z) T_{\rm cmb,0}$ (Black curve in Fig.~\ref{fig:EBL_CMB}), 
where $T_{\rm cmb,0}=2.725$ K is the temperature at $z=0$. The associated CMB mean 
density, average energy and mean energy density are respectively:
\begin{align} \label{CMB}
   n_{cmb} &= 16 \pi \zeta(3) \left( \frac{k_B T_{\rm cmb}}{hc} \right)^3
   \approx 411 (1+z)^3 \quad \rm{cm}^{-3}, \\
   \epsilon_{cmb} &= \frac{\pi^4}{30 \zeta(3)} k_B T_{\rm cmb}
   \approx 6.34 \times 10^{-4} (1+z) \quad \rm{eV}, \\
   \rho_{cmb} &= n_{cmb} \epsilon_{cmb} \approx 0.26 (1+z)^4\quad \rm{eV.cm}^{-3},
\end{align}
where $\zeta(3) \approx 1.202$, $k_B$, and $h$ are the Ap\'ery's, the Boltzmann's, and 
the Planck's constants respectively. In the Thomson regime, leptons of local energy 
$E_e$ up-scatter soft photons to typical energy:
\begin{equation} \label{eqn:Egamma}
   E_{\gamma} = \frac{4 \epsilon_{cmb} E_e^2}{3 m_e^2 c^4}
   \approx 3.23 (1+z) \left( \frac{E_e}{1\rm{TeV}} \right)^{2} \quad  \rm{GeV}.
\end{equation}
Between two Compton scatterings, the leptons travel a Compton mean free path 
$\lambda_{ic} = 1/(n_{cmb} \sigma_{T}) \approx 1.19 $ kpc (corresponding to a 
scattering time of $t_{ic}=\lambda_{ic}/c = 3870$ yr), where $\sigma_T$ is the 
Thomson cross section. And on average, they loose energy at rate:
\begin{equation}
   \frac{dE_e}{dt} = \frac{4}{3} c \sigma_T \rho_{cmb} \left( \frac{E_e}{m_e c^2} \right)^2.
   \label{eqn:dE_dt}
\end{equation}
After a flight of length $x$, the lepton energy is $E_e(x) = E_e^0 / (1+x/D_{ic}^0)$
where
\begin{equation} \label{eqn:Dic}
   D_{ic}^0 =  \frac{3 m^2 c^4}{4 \sigma_T \rho_{cmb} E_e^0} \approx
   367~ (1+z)^{-4} \left( \frac{E_e^0}{1\rm{TeV}} \right)^{-1} \quad \rm{kpc},
\end{equation}
is the initial Compton cooling distance, and $E_e^0$ is the lepton energy at the 
production site.

\subsection{Observables}
\label{sec:observables}
Based on these simple properties, some analytical estimates for typical cascade 
observables can be derived. Here we make the following standard assumptions:
\begin{itemize}
   \item The source of primary gamma-rays is isotropic and mono-energetic with an energy 
      $E_{\gamma}^0$.
   \item The primary gamma-rays all annihilate at exactly the same distance
      $\lambda_{\gamma\gamma}$ as given in Fig.~\ref{fig:lambda_gg}.
   \item The leptons are produced by photon-photon annihilation in the direction of the 
      parent gamma-ray photon and have exactly half of its energy.
   \item Compton interactions occur in the Thomson regime and the leptons travel 
      exactly over one mean free path $\lambda_{ic}$ between two Compton scatterings.
   \item The leptons are deflected by a uniform magnetic field perpendicular to the 
      motion. 
   \item Magnetic deflections occur locally on scales much smaller than the photon 
      annihilation length and the source distance.
   \item The scattered photons all get exactly the average energy given by equation
      \ref{eqn:Egamma} and are emitted in the propagation direction of the scattering 
      lepton.
   \item The absorption of these scattered photons (hereafter refereed to as first 
      generation photons) and the associated pair production is neglected.
      The contribution of higher generation particles is not considered.
   \item Cosmological effects are neglected.
\end{itemize}

With the above assumptions we can derive the following cascade geometrical and
distribution properties:

\subsubsection*{Geometry:}
Geometrical effects (halo effects and time delay) are due to the lepton deflection in the
extragalactic magnetic field. Two cases can be studied:
\begin{itemize}
   \item If the coherence length is large ($\lambda_B \gg D_{ic}^0$) the magnetic field 
      can be considered as uniform and the lepton deflection after a travel distance 
      $x$ is: 
      $\delta = \int_0^x ds/R_L(s)$. It means that a lepton of initial energy 
      $E_e^{0} = E_\gamma^0/2$ having cooled down to energy $E_e$ has been deflected 
      from its original trajectory by an angle:
      \begin{equation} 
         \delta = \frac{D_{ic}^0}{2 R_L^0} \left[ \left(\frac{E_e^0}{E_e}\right)^2-1 \right],
         \label{eqn:delta_ic0}
      \end{equation}
      where $R_L^0$ (Eq. \ref{eqn:RL}) is the initial Larmor radius of the lepton. 
      As soon as the lepton has lost a significant fraction of its energy, the previous 
      equation reduces to:
      \begin{equation} 
         \delta \approx \frac{D_{ic}}{2 R_L} = \frac{e_cB \lambda_{ic}}{2E_{\gamma}},
            \label{eqn:delta_ic}
      \end{equation}
      where $D_{ic}$ and $R_L$ are no longer the initial values but are now evaluated locally 
      at energy $E_e\ll E_e^0$, corresponding to photons scattered to energy $E_\gamma$
      (Eq. \ref{eqn:Egamma}).

   \item If the coherence length is short ($ \lambda_B \ll D_{ic}^0$), the leptons travel 
      across many zones of a highly turbulent field. We assume that the field is composed 
      of many cells of size $\lambda_{B}$, with uniform field and random directions. In each 
      cell, the leptons are deflected by an angle $\sim \lambda_{B}/R_{L}$ in a random 
      direction. Considering this process as a random walk leads to:
      \begin{align} 
         \delta & = \frac{\sqrt{D_{ic}^0 \lambda_{B}}}{2 R_L^0} \left[ 
                     \left(\frac{E_e^{0}}{E_e}\right)^2-1 \right] \nonumber \\
                & \approx \frac{\sqrt{D_{ic} \lambda_{B}}}{2 R_L}, 
               \label{eqn:delta_ic2}
      \end{align}
      where the last approximation is obtained similarly to Eq. \ref{eqn:delta_ic}.
\end{itemize}
In both cases, the magnetic deflection is a function of the lepton energy $\delta(E_e)$ 
and of the secondary gamma-ray energy $\delta(E_\gamma)$. In the following, we will 
concentrate on large-coherence case ($\lambda_B \gg D_{ic}^0$). However, similar 
constraints are easily obtained for short coherence lengths.

\begin{figure} \centering
   \includegraphics[width=\columnwidth]{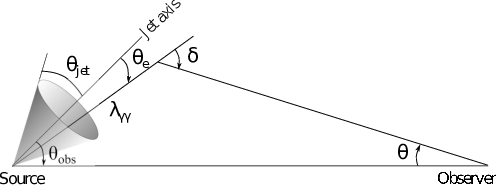}
   \caption{Geometry of the one-generation model. }
   \label{fig:triangle}
\end{figure}

The geometrical properties of the cascade (extension, time delay) can be derived from 
the magnetic deflection angle \cite[e.g.][]{neronov_method_2007,dermer_time_2011}. They 
are illustrated in Fig.~\ref{fig:triangle}. 
In the one-generation approximation, halo photons observed with a finite angle $\theta$ 
were emitted out of the line of sight and then deflected back to the observer. Assuming 
the lepton deflection occurs on very short distances compared to the photon absorption 
length, the detection angle and time delay are:
\begin{align}
   \theta &= \arcsin \left( \frac{\lambda_{\gamma \gamma}}{D_s} \sin\delta \right)
         \approx \frac{\lambda_{\gamma \gamma}}{D_s} \delta, \label{eqn:theta}  \\
   c \Delta t &= \lambda_{\gamma\gamma} (1- \cos \delta) - D_s (1- \cos \theta)
         \approx \lambda_{\gamma\gamma} \frac{\delta^2}{2}, \label{eqn:time_delay}
\end{align}
where $D_s$ is the distance to the source, $\lambda_{\gamma\gamma}$ is the annihilation 
distance of the primary photons, and where the last approximations were obtained in the 
small-angle approximation ($\lambda_{\gamma\gamma}\ll D_s, \delta \ll 1$). These 
relations are illustrated in Figs. \ref{fig:simple_case-observables} in the case of 
large coherence length.

As high-energy leptons travel and get deflected, the detection angle and 
time delay decrease as energy increases. In this regime, the small angle 
approximation is well satisfied and for large coherence length ($\lambda_B \gg D_{ic}^0$) 
the latter equations combined to Eq. \ref{eqn:Egamma} and \ref{eqn:delta_ic} yield 
respectively to:
\begin{align}
   \theta & \approx 0.79^\circ \left( \frac{\tau_{\gamma\gamma}}{397.4} \right)^{-1}
                           \left(\frac{B}{10^{-14}\,\textrm{G}}\right)
                           \left(\frac{E_\gamma}{1\,\textrm{GeV}}\right)^{-1}
                           \label{eqn:theta_approx}, \\
   \Delta t & \approx 65 \left(\frac{\lambda_{\gamma\gamma}}{1.32 \rm Mpc}\right)
                           \left(\frac{B}{10^{-17}\textrm{G}}\right)^2
                           \left(\frac{E_\gamma}{1\,\textrm{GeV}}\right)^{-2} \,\textrm{yrs}    
   \label{eqn:Dt_approx},
\end{align}
where $\tau_{\gamma\gamma} =D_s/ \lambda_{\gamma\gamma}$ is the annihilation optical 
depth. All values are calculated for a source at $z=0.13$ emitting primary photons at 
$E_\gamma^0=100$ TeV. These equations show the complementarity in the search for pair 
halos and pair echoes. Pair halos can only be observed if they are larger than the 
instrument point spread function (PSF). For a typical PSF of $0.1^\circ$, the above 
values correspond to magnetic fields larger than $10^{-14}$ G. Hence, large 
magnetic field strengths can be constrained through detection of pair halos. In this 
case, time delays are as long as $10^8$ yrs and echoes cannot be observed. On the 
other hand, still with the values above, observable echoes shorter than $5$ yr require magnetic
fields lower than $10^{-18}$-$10^{-17}$ G. Hence low magnetic field strengths can be constrained 
though detection of pair echoes. In that case, pair halos are typically smaller than 
$ 0.0001^\circ$ and cannot be resolved.

As the lepton energy decreases, both the detection angle and the time delay increase 
until the maximal deflection $\delta = \pi/2$ is reached for which $\theta_{\rm max} 
\approx 1/\tau_{\gamma\gamma} = 5.7^\circ (\tau_{\gamma\gamma}/10)^{-1}$. This 
corresponds to the maximal halo size. At a lower lepton energy, the Larmor radius becomes 
smaller than the magnetic coherence length. 
The leptons are trapped by the magnetic field and cannot travel farther.
This leads to the formation of a cloud of  $e^{+}-e^{-}$ pairs around the source. 
The size of the observed halo then corresponds to the physical extension of the pair 
cloud, i.e. $\lambda_{\gamma\gamma}$.

\subsubsection*{Distributions:}
In the following, unless otherwise specified, all distributions are normalised to one 
single primary photons emitted.

The cascade spectrum produced by one single high-energy photon is computed as $dN/dE_\gamma 
= 2 (dN/dt)_{ic} / (dE_\gamma/dE_e) / (dE_e/dt)$, where the factor 2 accounts for the two leptons produced by one single primary photon. Noting that the number of photons up-scattered by one lepton per unit time is $(dN/dt)_{ic} = c/\lambda_{ic}$, and using Eq. \ref{eqn:Egamma} and \ref{eqn:dE_dt} to 
derive the last terms leads to: 
\begin{equation} \label{eqn:spectrum}
   E_{\gamma}^2 \frac{dN}{dE_{\gamma}} =
   \frac{m_e c^2}{2} \sqrt{\frac{3 E_{\gamma}}{\epsilon_{cmb}}}
   \approx 556 \left( \frac{E_{\gamma}}{1\,{\rm GeV}}\right)^{1/2} (1+z)^{-1} \textrm{GeV}.
\end{equation}
It is a simple power-law $dN/dE_\gamma \propto E_{\gamma}^{-\Gamma}$ with index 
$\Gamma =3/2$. If unabsorbed, this spectrum extends up to the energy of photons 
scattered by the highest-energy leptons, i.e. 
$E_{\gamma,\rm max} = 0.8 \left(E_{\gamma,0}/1~{\rm TeV}\right)^2$~GeV. However,
for distant sources, such a power-law spectrum is typically cut at lower energies by
photon absorption (Fig.~\ref{fig:cutoff_energy}). In principle, a new generation of 
particles is then produced. Higher photons generation have often been neglected 
although they may contribute significantly to the overall spectrum (see section 
\ref{sec:simple_case}).

The angular distribution can be computed as $dN/d\theta = (dN/dE_\gamma)
(dE_\gamma/d\delta)(d\delta/d\theta)$. Using Eq. \ref{eqn:spectrum}, \ref{eqn:delta_ic}, 
and \ref{eqn:theta} to estimate the three factors respectively allows to derive 
the angular distribution of first-generation photons produced by one single primary photon, 
for large coherence length and small angle:
\begin{align}
  \theta \frac{dN}{d\theta}  &= m_e c^2 \left( \frac{3}{2  \lambda_{ic} \epsilon_{\rm cmb}}
      \frac{\tau_{\gamma\gamma} \theta}{  e_cB} \right)^{1/2} \nonumber \\
   &= 1972.3 \left( \frac{\tau_{\gamma\gamma}}{397.4} \right)^{1/2} \left( \frac{B}{10^{-15} \rm G}
      \right)^{-1/2} \left( \frac{\theta}{1^\circ} \right)^{1/2}. 
   \label{eq:theta_distrib}
\end{align}
Identically, writing $dN/dt = (dN/dE_\gamma)(dE_\gamma/d\delta)(d\delta/dt)$ and 
using Eq. \ref{eqn:time_delay} gives the following distribution of time
delays:
\begin{align}
   \Delta t \frac{dN}{dt}  &= m_e c^2 \left( \frac{3 }{ 4\lambda_{ic} 
      \epsilon_{cmb} e_c B} \right)^{1/2}  \left(\frac{c\Delta t}{2\lambda_{\gamma\gamma}}
      \right)^{1/4} \nonumber \\
   & =  437.2 \left(\frac{\lambda_{\gamma\gamma}}{1.32 \rm Mpc} \right)^{-1/4}
      \left(\frac{\Delta t}{10^6 \rm yr} \right)^{1/4}  \left(\frac{B}{10^{-15} \rm G}
      \right)^{-1/2}. 
   \label{eq:Dt_distrib}
 \end{align}
These distributions are pure power-laws and do not show any specific angular (or time) scale
because they are composed by the contributions of photons detected with all possible energies. 
However, as discussed below, real observations are obtained in a limited energy range, with a 
given aperture angle, and with a finite observation duration. This produces characteristic scales 
in energy, arrival angle and time delay, that appear as cuts or breaks in the corresponding 
distributions. 

\section{Code description and test cases}
\label{sec:simple_case}

The analytical approximations presented in the previous sections are useful to understand the
physics of the cascades and to obtain orders of magnitude estimates, but accurate calculations 
require numerical simulations. For this purpose, we have developed a new Monte Carlo code. In this
section, we outline the main features of this code and compare its results to the analytical
predictions presented in section \ref{sec:observables} as well as to other published results.

\subsection{Code and simulation set-up}\label{sec:code}

Our Monte Carlo code is  designed to track the propagation of all particles and to reproduce the
properties of the cascade as precisely as possible. A complete description of the code can be 
found in appendix \ref{appendix:code}. However, the main algorithm is summarised here (see also 
Fig.~\ref{simulation_process}):
\begin{itemize}
   \item Primary particles (photons or leptons) are launched at a given redshift and with a given 
      energy. Particle energy can also be sampled from a power-law distribution. In this
      paper will focus only on cascades initiated by photons.
   \item Interaction distances are generated randomly according to the probability distribution
      given by the exact cross-sections ($e^{+}-e^{-}$ pair production for a photon, Klein-Nishima
      cross-section for a lepton) and taking into account the cosmological evolution of the target
      particles.
   \item The particles are propagated taking into account all cosmological effects (redshift,
      expansion). In particular, the transport of leptons in the EGMF is computed as described in
      appendix \ref{appendix:travel}. If the particles interact before reaching the Earth, the 
      interaction outcome is computed. Energy and direction of the outcoming particles are 
      generated according to the probability distributions given by the exact differential
      cross-sections. 
      Particles with updated parameters and new particles are then stored for later treatment.
      The physical parameters of the particles reaching the Earth without any interaction
      are stored for post-processing.
      
   \item The extragalactic magnetic field is modeled by dividing the comoving space into a number 
      of cells with size $\lambda_B$, defining regions of uniform magnetic field. 
      For each cube a random magnetic field direction is computed and is kept for the entire 
      simulation. The EGMF is set by its strength at redshift $z=0$. 
      For very short coherence lengths, the lepton interaction distance ($\lambda_{IC}\approx 1$ 
      kpc) can become larger than $\lambda_B$. In such a case, the motion to the next interaction 
      location is divided in shorter steps of length a fraction of $\lambda_B$, to ensure
      that the leptons are deflected by all cells. 
\end{itemize}

The following subsections present the results of test simulations. For the sake of 
comparison, we first consider a simple canonical model consisting in a mono-energetic source 
(100 TeV) emitting isotropically at z=0.13 (around 557 Mpc). We use the EBL model of 
\cite{dominguez_extragalactic_2011}. With this set-up the characteristic annihilation distance 
of primary photons is $\lambda_{\gamma\gamma}=1.32$ Mpc. The 
EGMF is set to $B=3 \times 10^{-16}$ G with a coherence length of $\lambda_B=1$ Mpc. 

We focus on the three main observables characterising the photons detected on Earth: their 
energy, their arrival angle, and their time delay. We derive the 3D photon distribution in 
this space of parameter. We also concentrate on the contribution of the different 
{\it generations} of particles, i.e. their rank in the cascade tree. In the following, the 
zeroth generation corresponds to the primary photons emitted at the source. These photons 
annihilate, produce pairs, that in turn produce the photons of first generation. Again, these 
can produce pairs which up-scatter target photons producing the second photon generation, and 
so on. As we are only interested in the detection of photons, their energy will be written $E$, 
where the subscript $\gamma$ has been dropped for simplicity.

\subsection{Correlations between observables}
\label{sec:correlation}

\begin{figure}
   \centering
   \includegraphics[width=\columnwidth]{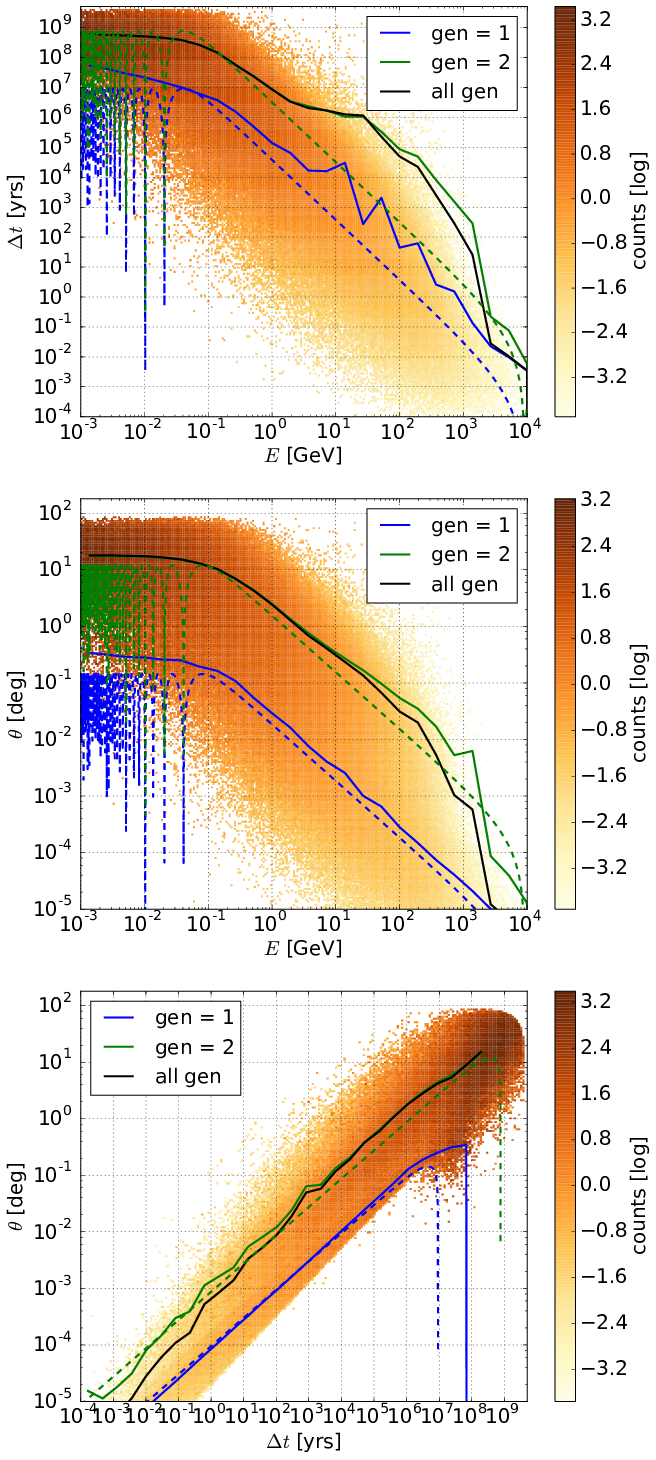} 
   \caption{{\bf Top panel:} Correlation between the time delay $\Delta t$ and energy $E$ of 
      detected photons. The density map shows the number of photons per unit energy and time 
      delay $(E \Delta t) d^2N/(dE d\Delta t)$ with a log color scale. The blue, green and 
      black solid lines show the average time delay for first-generation photons only, 
      second-generation photons, and all photons, respectively. The blue and green dashed lines 
      show the analytical estimates for the first and second generations only (see sec. 
      \ref{sec:observables}). With the same notations, the {\bf middle and bottom panels} show 
      the energy-detection angle and detection angle-time delay correlations respectively. }
   \label{fig:simple_case-observables}
\end{figure}

In the framework of the approximations used in the one-generation analytical model, the photon 
energy, detection angle and time delay are linked exactly through simple relations. When these 
approximations are relaxed a significant scatter is expected. Fig. 
\ref{fig:simple_case-observables} show the three possible correlations between 
the photon energy $E$, the detection angle $\theta$, and the time delay $\Delta t$. 
In addition, the {\it average} behaviour is plotted in solid line for each generation by binning
the x-axis (with 4 bins per decade) and averaging the y-values. The analytical estimates of the
one-generation model (from Eq. \ref{eqn:theta} and \ref{eqn:time_delay}) are shown for comparison 
as blue, dashed lines. The expected trends are recovered: the smaller the energy, the larger the 
observation angle and the arrival time delay. 
As the leptons cool down, they produce more and more gamma-ray photons per unit time. As a
result, the total cascade emission is dominated by a very large number of photons with low-energy, 
large angle, long time delay, as can be seen in Fig.~\ref{fig:simple_case-observables}.
However, several important points can be made.

The averaged values for the first-generation photons (blue line) are consistent with the
analytical results of sec. \ref{sec:observables}: both the saturation at low energy and the slopes 
in the small-angle regime are well recovered. The saturations observed in time correspond to 
the maximal delay (photons that are emitted away from, and scattered back to the observer) while 
the saturation observed in detection angle corresponds to photons deflected by an angle of 
$\delta=\pi/2$. The power-law regimes correspond to Eq. \ref{eqn:theta_approx} and
\ref{eqn:Dt_approx}. Most of the observed deviations at high energy (e.g. fluctuations and 
peaks) come from the limited statistics of the simulations: averaged values can be contaminated 
by a few photons with very large values (typically photons scattered by EBL targets instead of CMB 
photons). Physical processes that are not taken into account in the analytical estimates 
(Klein-Nishina regime, dispersion in the annihilation distance, non-uniform magnetic field, energy 
dispersion around the averaged value \dots) are also responsible for deviations to the 
approximations, but the effects are weaker.

These results also show that the one-generation model underestimates the detection angle and the
time delay by at least two orders of magnitude when the contribution of second-generation photons
is significant. Indeed the highest energy, second-generation photons are almost all produced 
at the location where the parent leptons which emitted them where produced. As these photons have 
lower energy than the primary photons, their annihilation distance is larger 
$\lambda_{\gamma\gamma}^{gen=1} \gg \lambda_{\gamma\gamma}^{gen=0}$. As a result, the highest 
energy, second-generation photons are typically produced at a distance 
$\lambda_{\gamma\gamma}^{gen=1}$ from the source. The geometry remains however similar so that an 
estimate for the second-generation observable quantities can be found by substituting 
$\lambda_{\gamma\gamma}$ by $\lambda_{\gamma\gamma}^{gen=1}$ in the results of sec. 
\ref{sec:observables}. For primary photons at 100 TeV, the highest-energy, first-generation 
photons ($8$ TeV) have mean free path $\lambda_{\gamma\gamma}^{gen=1}=117$ Mpc. This is shown 
in Fig.~\ref{fig:simple_case-observables} as green dashed lines. As can be seen, this new estimate 
matches well the average results of second-generation dominated cascades.

We emphasise that a single generation never dominates at all energies, angles and time delays. 
In our canonical simulation for instance, only the low-energy spectrum is dominated by 
second-generation photons while the highest energy photons are mostly first-generation photons 
(see Fig.~\ref{fig:simple_case-spectrum}). This is why the average time delay drops below
the second-generation estimates at high energy. 

In principle, the ratio of first to second generation photons depends on the energy 
of primary photons and the source distance. If primary photons have energy smaller than or close 
to the absorption energy $E_{\rm abs}$, then the absorption is so weak that the production 
of second-generation photons is quenched, and the cascade is dominated by first-generation photons 
(this will be illustrated in Fig. \ref{fig:spectrum_vs_powerlaw} for instance). However, the 
results presented here remain general as soon as the energy of primary photons is significantly 
larger than the absorption energy \citep[see also][]{berezinsky_high-energy_2016}.

In any case, there is a large dispersion around the average quantities, showing that the latter
might be of limited practical use.

\subsection{Photon distributions}

The energy, angle and time distributions are obtained by integrating over the other two
parameters. In this section we focus on photons with energy above 1 GeV which corresponds to 
the typical energy range of gamma-ray observatories.

\begin{figure} \centering
   \includegraphics[width=\columnwidth]{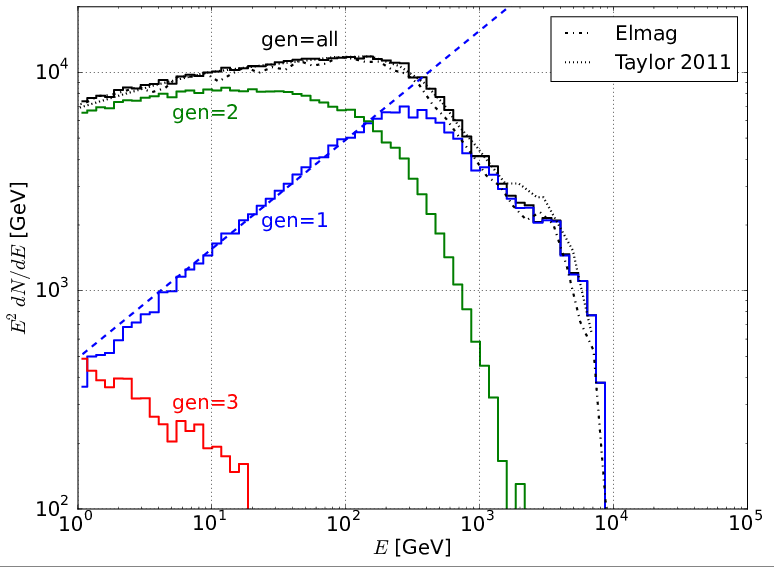}
   \caption{Full spectrum for an isotropic and mono-energetic source (100~TeV) 
   at $z$=0.13 (black line). The blue, green and red lines show the contributions of 
      generations 1, 2 and 3 respectively. The analytic expression (Eq.~\ref{eqn:spectrum}) is 
      shown in blue dashed line. The dotted-dashed, and dotted lines show the results from Elmag
      \citep{kachelriess_elmag_2012} and \citet{taylor_extragalactic_2011} respectively. }
      \label{fig:simple_case-spectrum}
\end{figure}

The spectrum of our fiducial model, integrated over the entire sky, over all possible arrival
times, and normalised to one primary photon, is shown in Fig.~\ref{fig:simple_case-spectrum}. 
It is compared to Fig.~1 of \citet{taylor_extragalactic_2011}\footnote{The normalisations of the 
published results where chosen arbitrary} and to the results of the Elmag code 
\citep{kachelriess_elmag_2012} using the same set-up. Spectral shapes provided by the different 
codes are compatible each other.
All primary photons at 100 TeV are absorbed and produce the first-generation spectrum (blue 
line). Below $E_{\rm cut}$ analytical estimates (Eq. \ref{eqn:spectrum}) are well reproduced 
by the first-generation population, in spite of the approximations made. Above 1 TeV the photon 
of the first generation are
absorbed (Fig.~\ref{fig:cutoff_energy}) and produce a second generation which dominates
the spectrum below about 100 GeV (green line). The spectrum of the second generation is 
similar to the spectrum of the first generation ($dN/dE\propto E^{-3/2}$) except it is softer 
in the energy range shown in this figure (it can be shown that $dN/dE \propto E^{-7/4}$). A few 
second-generation photons are also absorbed and produce a weak third-generation population 
(red line) which does not contribute to the total spectrum.

\begin{figure} \centering
   \includegraphics[width=\columnwidth]{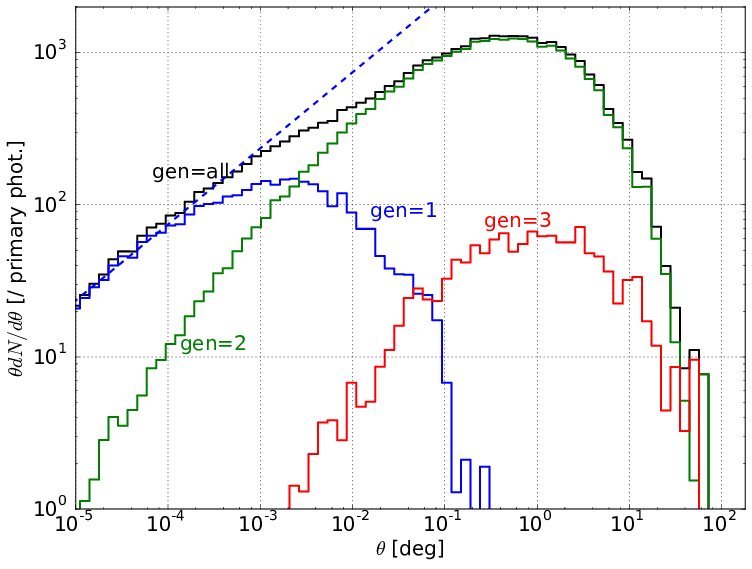}
   \caption{Detection angle distribution for $E>1$ GeV photons. The colors are the same as in Fig.
      \ref{fig:simple_case-spectrum}. The blue, dashed line shows the analytical estimate of 
      Eq. \ref{eq:theta_distrib}. }
   \label{fig:simple_case-arrival_angle_distribution}
\end{figure}

The angle distribution integrated over energies $E>1$ GeV
and over all arrival times, normalised to one primary photon is shown in 
Fig.~\ref{fig:simple_case-arrival_angle_distribution}.
The emission is peaked at the center and decreases as a power-law with increasing angle. At small 
angles, the distribution of first-generation photons is well approximated by the analytical 
estimate given in Eq. \ref{eq:theta_distrib}: $dN/d\theta \propto \theta^{-1/2}$. However, 
only photons above 1 GeV are considered here. Hence, low-energy photons with large angles are not 
observed. And, as compared to the analytical estimate, the  angular distribution drops at a 
typical size that depends on the minimal energy and the magnetic field. As second-generation 
photons have a larger mean free-path, they typically arrive with larger angles. Interestingly, the 
distribution is dominated by second-generation photons at observable scales ($\theta>0.1^\circ$). 
This result remains general as long as the energy of primary photons is significantly larger than 
the absorption energy $E_{\rm abs}$ (i.e. photons have absorption depth $\tau_{\gamma\gamma}(E^0)>>1$). 
Primary photons with energy comparable to the absorption energy  ($\tau_{\gamma\gamma}(E^0)
\approx 1$) are weakly absorbed and do not produce high-generation photons. For sources with an extended intrinsic spectrum, the contribution of first- and second-generation photons to the angular distribution is more complex (see sec. \ref{sec:source_spectrum}).

\begin{figure} \centering
   \includegraphics[width=\columnwidth]{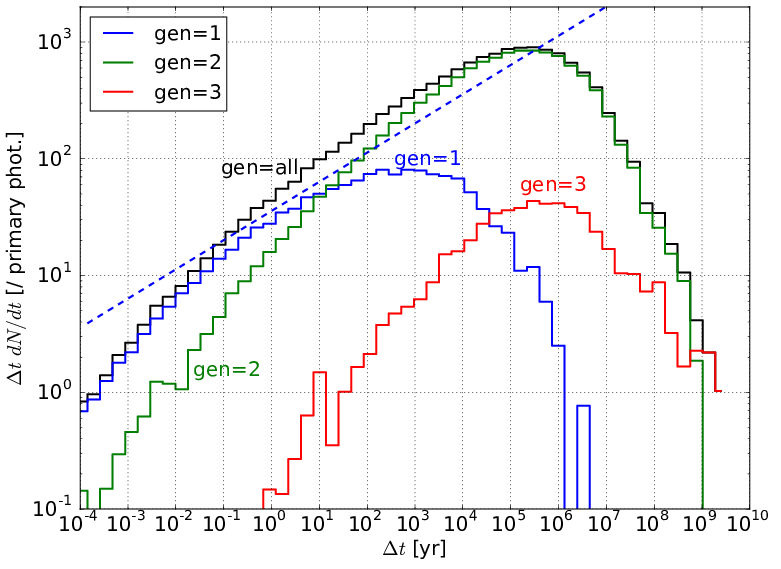}
   \caption{Distribution of time delays for $E>1$ GeV photons. The colors are the same as in Fig.
      \ref{fig:simple_case-spectrum}. The blue, dashed line shows the analytical estimate of 
      Eq. \ref{eq:Dt_distrib}. }
   \label{fig:simple_case-time_delay_distribution}
\end{figure}

The time delay distribution integrated over energies $E>1$ GeV
and all detection angles, and normalised to one primary photon, is shown in Fig.
\ref{fig:simple_case-time_delay_distribution}.
As time evolves after a source flare, less and less photons are observed. The one-generation model 
(Eq. \ref{eq:Dt_distrib}) provides a good estimate of the first-generation distribution for time 
delays ranging from about 1 month to 100 years. Shorter time delays correspond to high energy 
photons that are absorbed, producing a drop below the analytical estimate. Longer time delays 
correspond to low-energy photons below the selection criterion $E>1$ GeV, producing a cut of 
the distribution above 1000 years. Interestingly, accessible time delays ($\Delta t < 1$ yr) 
after a flaring event (such as a GRB) are short enough to be dominated by first-generation 
photons only, allowing for the use of simple formulae to derive constraints from the potential 
detection of pair echoes.

\begin{figure} \centering
   \includegraphics[width=\columnwidth]{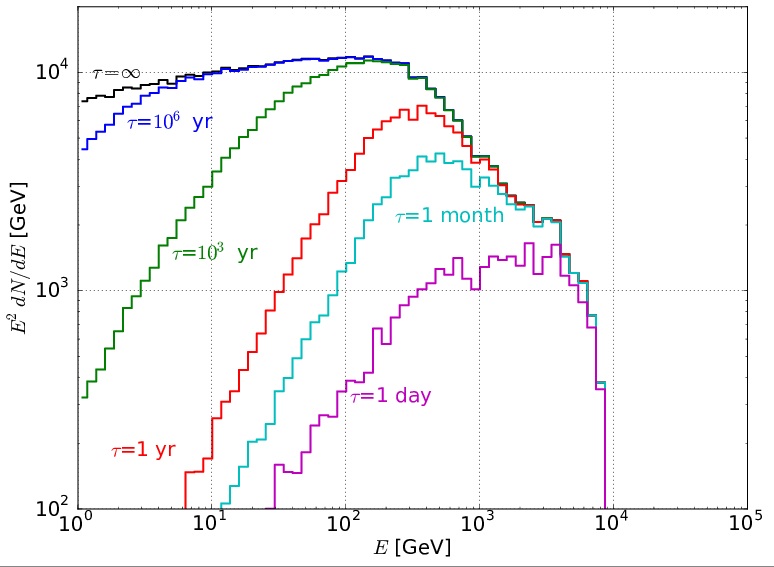}
   \caption{Total spectrum of an flaring event integrated over a finite exposure time 
      ($t_{\rm obs}=\tau$), or equivalently: instantaneous spectrum of a source that has 
      been active for a give time ($t_{\rm act}=\tau$) in the past.}
   \label{fig:simple_case-spectrum_tobs}
\end{figure}

The three distributions presented on Figs.~\ref{fig:simple_case-spectrum},
\ref{fig:simple_case-arrival_angle_distribution} and
\ref{fig:simple_case-time_delay_distribution} ($dN/dE$, $dN/d\theta$, and $dN/dt$) are
global distributions in the sense that they are integrated over large ranges of the 2 others
quantities. For instance, the spectrum shown in Fig.~\ref{fig:simple_case-spectrum} is the
integrated over all detection angles and all arrival times. However, in a more realistic
situation, only limited ranges of these quantities are accessible. As low-energy photons arrive
with large angle, large time delay, and are products of primary photons emitted far away from the
line of sight, any of the following effects will damp the spectrum at low energy, while cutting
the large angle and large time delay part of the associated distributions:
\begin{enumerate}
\item If the instrument is only sensitive above a given energy (see discussion before).
\item If the instrument aperture is limited.
\item If the exposure time is finite after an impulsive flaring event.
\item If the source emission is beamed within a limited opening angle.
\end{enumerate}
The effect of finite exposure time is  illustrated in Fig.~\ref{fig:simple_case-spectrum_tobs}
\citep[see also][]{ichiki_probing_2008}.
This can be interpreted in two ways. If the source produces a strong, impulsive flaring event (such
as a GRB for instance), this figure shows the integrated spectra as data is accumulated from the
detection of the unabsorbed, primary photons up to time $t_{\rm obs}=\tau$. As time evolves lower 
energy photons are detected and the low-energy part of the spectrum builds up slowly. 
Alternatively, this figure also shows the instantaneous spectra observed at present time, if 
the source (such as an AGN) has been active for a time $t_{act}=\tau$ in the past, with constant 
luminosity \citep{dermer_time_2011}. As the activity period increases, we are able to detect 
secondaries produced by primaries emitted earlier. As leptons had more time to cool down, 
these secondaries have lower energy. As a result, long activity sources have spectra that extend 
to lower energy.

\section{Source properties}
\label{sec:source_profile}

The simple case presented in section~\ref{sec:simple_case} allows us to understand general 
behavior of electromagnetic cascades. However, several effects must be included to produce 
realistic cascades. Gamma-ray sources (AGNs, GRBs) do not emit photons at a single energy
or isotropically but instead produce non-thermal, beamed radiation. Here we investigate the 
following intrinsic properties of the source:
\begin{itemize}
   \item Redshift $z$.
   \item Intrinsic spectrum:  here we consider power-law spectra in the form $dN/dE \propto
      E^{-\Gamma}$ for $100$ MeV$ \ge E \ge E_{\rm max}$.
   \item Emission profile: here we assume a disk emission, i.e. an axisymmetric angular
      distribution $dN_e/d\Omega_e$ uniform up to a given half-opening angle $\theta_{\rm jet}$
      observed with an angle $\theta_{\rm obs}$ away from its axis (see Fig. \ref{fig:triangle}).
\end{itemize}
In the following we use simulation parameters corresponding to the Blazar 1ES0229+200 
\citep{tavecchio_hard_2009,taylor_extragalactic_2011,vovk_fermi/lat_2012} with a
redshift $z=0.14$ corresponding to a distance of 599 Mpc, and a hard spectrum with $\Gamma=1.2$. 
The unconstrained maximal energy of the intrinsic spectrum is set to $E_{\rm max}=100$ TeV and the
emission is assumed to be isotropic. The EGMF has an averaged intensity of $B=10^{-15}$ G, and a 
coherence length $\lambda_B=1$ Mpc. We use the EBL model of \cite{dominguez_extragalactic_2011}.

In this section, the spectra are normalised to $L_0$, the intrinsic luminosity of the
source as observed at $z=0$ (i.e. the intrinsic luminosity decreased by a factor $1+z$).

\subsection{Source redshift}

\begin{figure} \centering
   \includegraphics[width=\columnwidth]{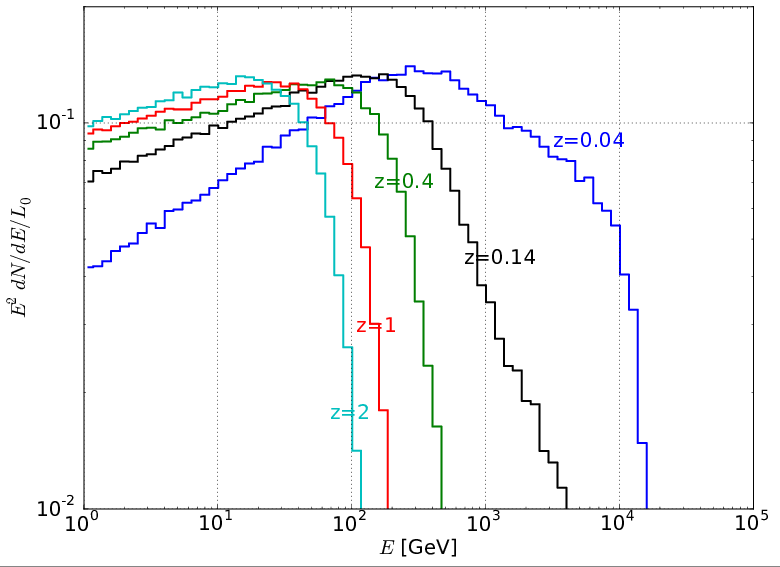}
   \caption{Full spectrum of sources at different redshifts ($z=0.04$,
      $0.14$, $0.4$, $1$ and $2$), normalised to the intrinsic luminosity attenuated by a factor 
      $1+z$ to account for the universe expansion.}
   \label{fig:spectrum_vs_redshift}
\end{figure}

The spectral evolution with redshift is shown in Fig.~\ref{fig:spectrum_vs_redshift} for 
$z=0.04$ ($D_s\sim 175$~Mpc), $z=0.14$ ($D_s\sim 599$~Mpc), $z=1$ ($D_s\sim 3.4$~Gpc), and $z=2$ 
($D_s\sim5.3$~Gpc).
As expected, the cut-off energy decreases with increasing distance as the column density of 
target photons increases. At high redshift, we find that the absorption depth goes simply as 
$\tau_{\gamma\gamma}\propto E^2$ producing a super-exponential cutoff $\propto 
e^{-E^2/E_{\rm cut}^2}$. However the spectra of nearby sources show a more complex and harder 
absorption cutoff. In our setup, the maximal energy of primary photons (100 TeV) is large enough 
to generate an efficient cascade. As a result, almost all shown spectra are dominated by 
second-generation photons. Only when the source is close enough ($z=0.04$), the absorption is 
weak enough to quench the production of second-generation photons. As discussed by 
\cite{berezinskii_cosmic_1975} the spectrum softens as the generation order increases, which is 
consistent with the harder spectrum observed at $z=0.04$. The intrinsic spectrum used here is hard 
($\Gamma=1.2$), so that most of the intrinsic luminosity is concentrated at the highest energies 
($E_{0} \sim E_{\rm max}$). Most of the spectrum is then fully absorbed and redistributed as 
cascade contribution. As a result, the amplitude of the observed spectra is almost insensitive 
to the absorption energy, i.e. also to the source redshift.

\begin{figure}
   \includegraphics[width=\columnwidth]{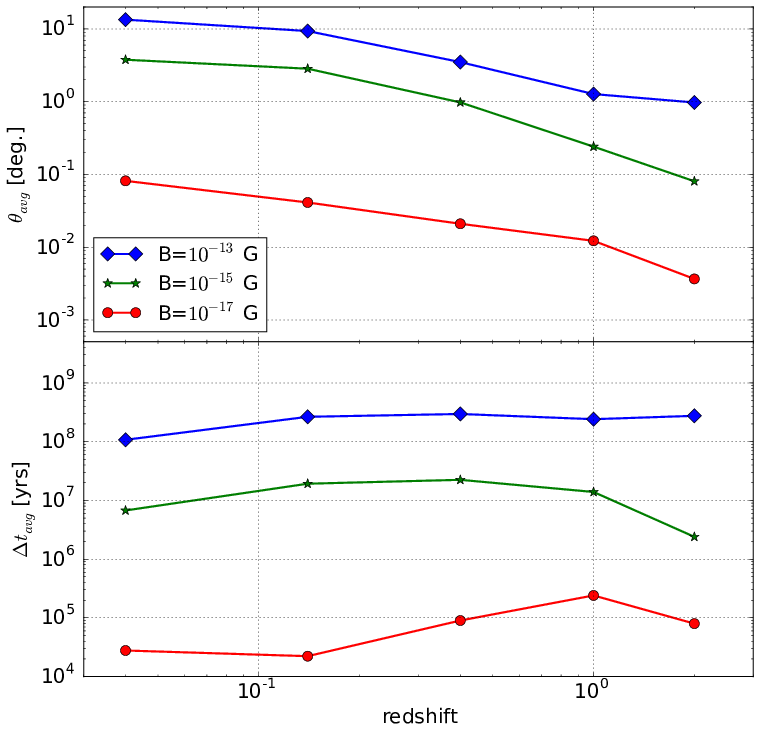}
   \caption{Average arrival angle (top) and time delay (bottom) of photons with energy $E > 1$ 
      GeV, as a function of the source redshift, for different EGMF strengths ($B=10^{-13}, 
      10^{-15}$, and $10^{-17}$ G). }
   \label{fig:mean_Dt_theta_vs_redshift}
\end{figure}

The evolution of the angular distribution and time delays of photons with energy $E > 1$ GeV, are 
illustrated by their average values in Fig.~\ref{fig:mean_Dt_theta_vs_redshift} for different EGMF 
strengths.
Both the halo extension and the time delay increase with magnetic field, as leptons of given 
energy are more deflected by stronger fields. Their evolution with redshift is the result of 
several effects. The halo extension decreases with distance. To zeroth order, it is simply due 
to geometrical effects: the same annihilation distance to the source corresponds to a smaller 
angle as seen from a more distant observer (see for instance Eq. \ref{eqn:theta} for 
first-generation photons: $\theta\propto \lambda_{\gamma\gamma}/D_s$). In contrast, the time 
delay does not suffer from any geometrical dependence on distance (see for instance Eq. 
\ref{eqn:time_delay} for first-generation photons) and shows only little evolution with 
redshift. To first order however, the cosmological evolution of the universe also influences 
the angular size and time delay of secondary photons (namely through $\lambda_{\gamma\gamma}(z)$, 
$\lambda_{ic}(z)$, $B(z)$, and $E(z)$), explaining the remaining evolution. 

\subsection{Source spectrum}
\label{sec:source_spectrum}

\begin{figure} \centering
   \includegraphics[width=\columnwidth]{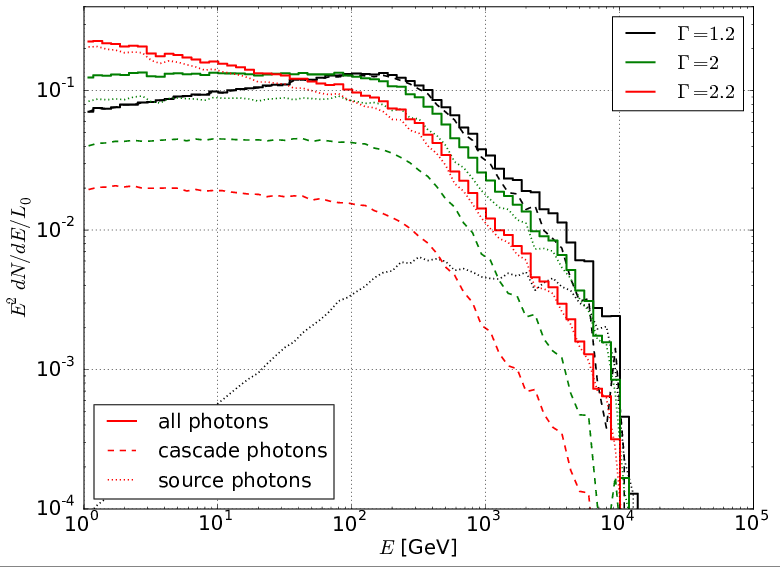} \\
   \includegraphics[width=\columnwidth]{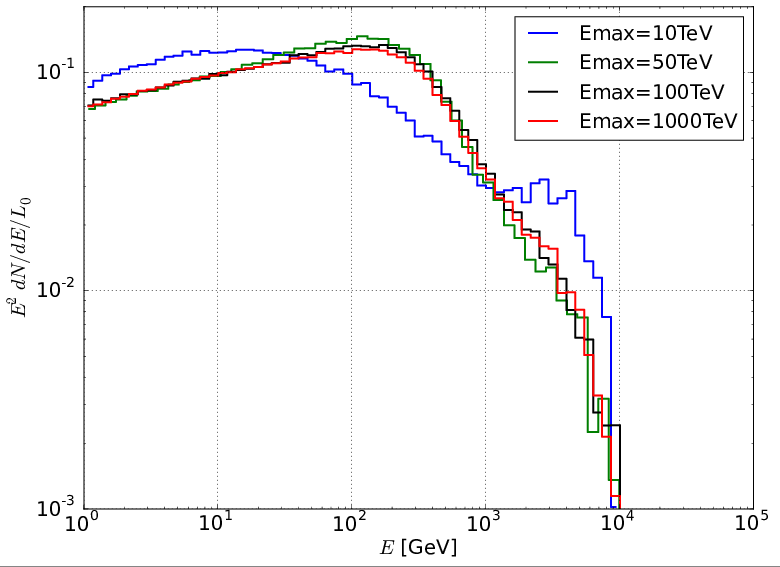}
   \caption{{\bf Top panel:} Observed spectra for $E_{\rm max}=100$ TeV and different spectral 
      indices  $\Gamma$ (solid lines) . The contributions of the primary 
      source and the cascade are shown in dotted and dashed lines respectively. Spectra are 
      normalised to $L_0$, the intrinsic luminosity attenuated by $1+z$. {\bf Bottom panel:} 
      Spectra for $\Gamma=1.2$ and different maximal energies $E_{\rm max}$.} 
   \label{fig:spectrum_vs_powerlaw}
\end{figure}

Fig.~\ref{fig:spectrum_vs_powerlaw} shows the observed spectra when the source intrinsic spectrum 
is changed.
The top and bottom panel show the results for different spectral indices ($\Gamma$=1.2, 
2 and 2.2, $E_{\rm max}=100$ TeV) and different maximal energies ($E_{\rm max}=10$, 50, 100 TeV 
and 1 PeV, $\Gamma=1.2$) respectively. 
At source distance $z=0.14$, photons with energy higher than a few TeV are absorbed 
and redistributed towards low energies. For hard spectra ($\Gamma<2$), many primary photons are 
absorbed. This induces a strong cascade which dominates the intrinsic source emission 
at all energies. The observed spectrum is very similar to the mono-energetic model shown in
Fig.~\ref{fig:simple_case-spectrum}. In contrast, when  the intrinsic spectra are soft 
($\Gamma>2$), only few primary photons are absorbed. The cascade contribution is negligible and 
only the absorbed, intrinsic spectrum is observed. As far as only the cascade emission is 
concerned (dashed lines), the spectrum is only weakly dependent on the intrinsic hardness. This 
strong universality of the cascade emission is also illustrated in the bottom panel where the 
shape of the observed spectrum is highly insensitive to the maximal energy, Only the case
with $E_{\rm max}=10$ TeV shows significant departure from the generic spectral shape. In that 
specific case, the intrinsic spectrum does not extend much beyond the absorption energy
$E_{\rm abs}$ at that distance, and the spectrum is dominated by first-generation photons which 
were produced below the average annihilation distance. 

\begin{figure} \centering
   \includegraphics[width=\columnwidth]{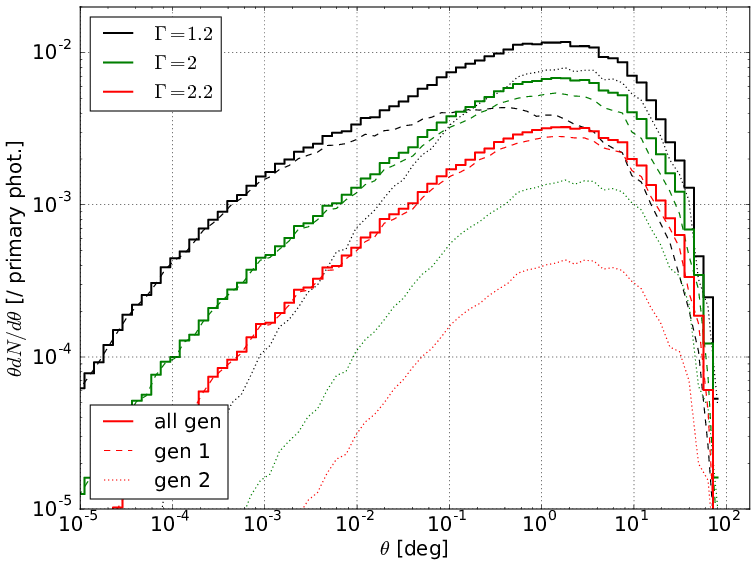}
   \caption{Angular distribution of $E>1$ GeV photons, for different intrinsic spectral indices 
      $\Gamma$ and $E_{\rm max}=100$ TeV (solid lines). The dashed and dotted lines show the 
      contributions of first-generation photons and second-generation photons respectively.} 
   \label{fig:dNdtheta_vs_powerlaw}
\end{figure}
Fig.~\ref{fig:dNdtheta_vs_powerlaw} shows the angular distribution in the 1 GeV-1 TeV band,
for different intrinsic spectral indices. As can be seen, the angular distribution is
rather insensitive to the intrinsic spectrum. 
Although the angular distribution of cascades induced by one single primary photon depends 
on the energy of this primary (it scales as $\lambda_{\gamma\gamma}(E_0)$, see Eq. 
\ref{eqn:theta_approx}), two effects contribute to keep the final distribution induced by 
an extended intrinsic spectrum quite universal. First, most of the cascade is induced by primary 
photons just above the absorption threshold (see Fig. \ref{fig:cutoff_energy}), i.e. in the 
1-10 TeV range, because they are more numerous. In this range the mean absorption length is rather insensitive to the primary
energy (see Fig. \ref{fig:lambda_gg}). Second, when the intrinsic spectrum extends to higher energy and 
is hard enough ($\Gamma \lesssim 2$), part of the cascade is in principle also induced by these high-energy
photons with shorter absorption length. However the first-generation photons produced by these
high-energy primaries are also absorbed. They produce second-generation photons that eventually dominate 
the angular distribution and have larger absorption depth (see sec. \ref{sec:correlation}). 
Here also, the second-generation photons contribute dominantly to the cascade when they have energy
just above the absorption energy, that is when they have absorption length comparable to that of
low-energy primaries. This makes the spectrum insensitive to the energy of primary photons, that
is to the intrinsic spectrum.
The time delay distribution is even less sensitive to the properties of the intrinsic spectrum, 
and is not shown here.

These results show that the emission properties of cascades (spectrum, angular distribution and 
time distribution) do not depend on the properties of the intrinsic spectrum as long as the 
maximal energy of the source is large enough compared to the absorption energy. Although the 
cascade properties are highly dependent on several other parameters such as the distance and the 
EGMF, this illustrates the universal properties of the cascade emission with respect to the source 
intrinsic spectrum \citep[e.g.][]{berezinsky_high-energy_2016}. 

\subsection{Source beaming}

\begin{figure} \centering
   \includegraphics[width=\columnwidth]{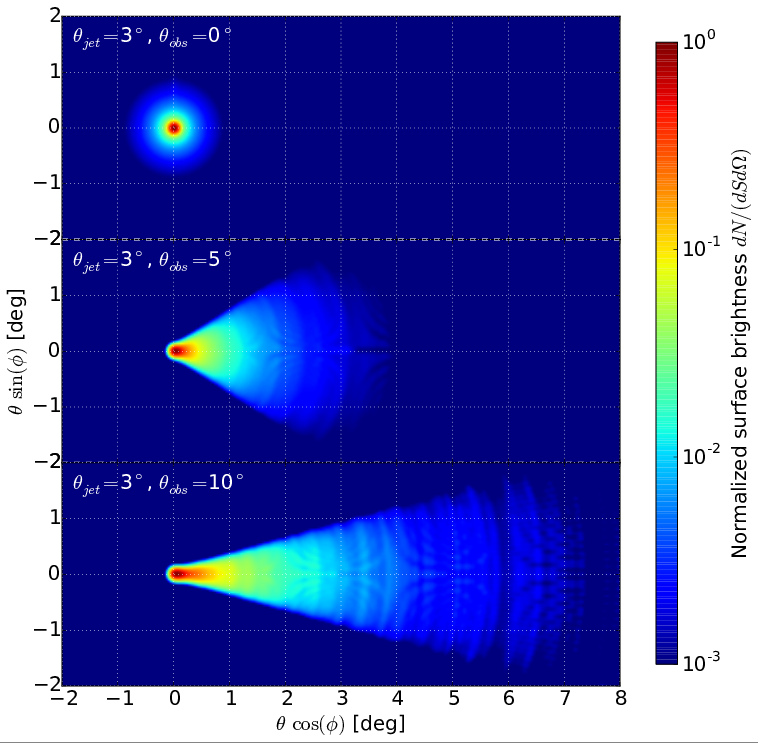}
   \caption{Images of a misaligned source at $z = 0.14$, emitting a power-law spectrum with index 
      $\Gamma =1.2$ in a uniform cone of half-opening angle $\theta_{\rm jet}=3^\circ$. Here 
      $B=10^{-15}$ G, and $\lambda_B=1$ Mpc. Observations at three angles at shown: 
      $\theta_{\rm obs}=0^\circ$, $5^\circ$, $10^\circ$ from top to bottom. }
   \label{fig:image_tobs}
\end{figure}

Anisotropic emission can be modelled at the post-processing stage if the source emission is 
axisymmetric (e.g. a beamed emission) and the intergalactic medium is isotropic (see appendix 
\ref{appendix:anisotropy}). If the source axis is not aligned with the line of sight, a jet 
structure is observed in the cascade emission \citep{neronov_degree-scale_2010,
arlen_intergalactic_2014}. This is illustrated in Fig.~\ref{fig:image_tobs}.
The general analysis of the misaligned case is beyond the scope of the current paper. In
the following, we shall concentrate on the case of a beamed emission aligned with the line 
of sight. 

\begin{figure} \centering
   \includegraphics[width=\columnwidth]{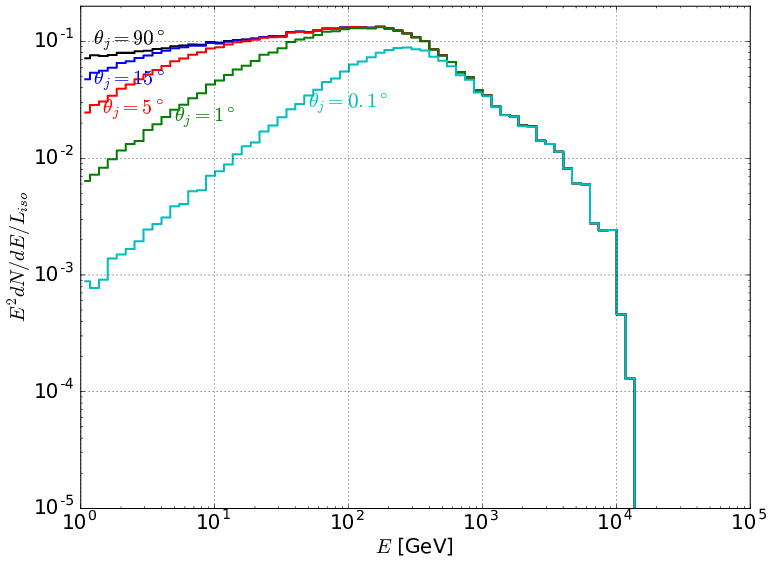}
   \caption{Energy spectra  for different jet half-opening angles $\theta_{\rm jet}$. All spectra 
      are normalised to the luminosity of an isotropic source of equal flux inside the emission 
      cone: $L_{\rm iso} = 2L_0/(1-\cos\theta_{\rm jet})$. }
   \label{fig:spectrum_vs_tjet}
\end{figure}

The effect of the jet half-opening angle $\theta_{\rm jet}$ on the observed spectrum is shown in 
Fig.~\ref{fig:spectrum_vs_tjet}.
Lower energy photons typically originate from primary photons emitted farther away from the 
line of sight. Therefore in the case of a beamed source with no photon emitted at large angle, 
the cascade emission is suppressed below some critical energy  
\citep[see for instance][]{tavecchio_intergalactic_2010}. Then, the more 
collimated the jet, the larger the critical energy. The precise transition energy depends on the 
magnetic field strength and coherence length. Although its value can be derived in the framework 
of a one-generation model, such results do not apply to sources emitting at high energy and 
producing high-generation dominated cascades. 

\begin{figure} \centering
   \includegraphics[width=\columnwidth]{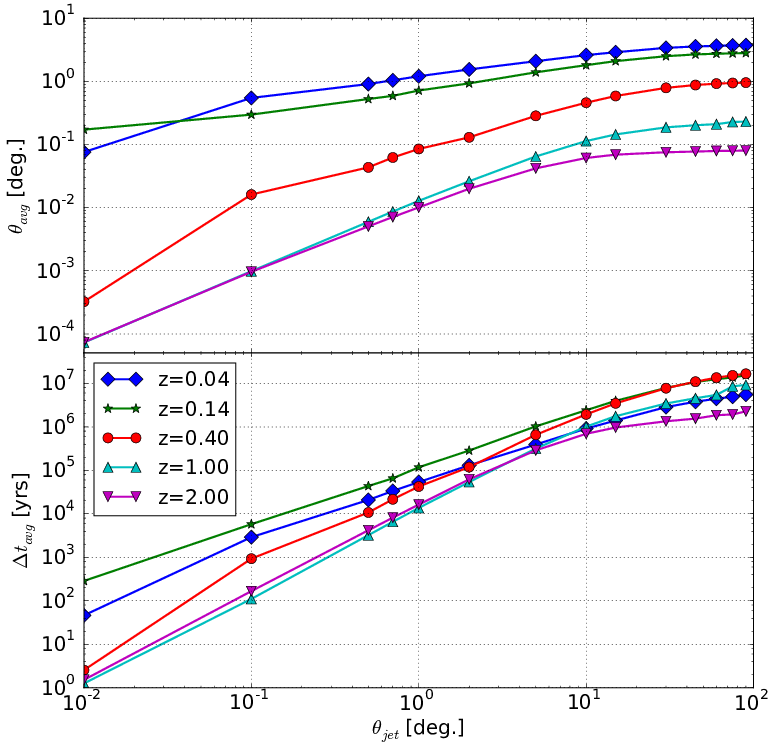}
   \caption{Average detection angle (top) and time delay (bottom) of photons with energy $E > 1$ 
      GeV as a function of the jet opening angle for sources at different redshifts. }
   \label{fig:meantheta_vs_jet_opening_angle}
\end{figure}

The effect of the jet opening angle on the observed angular distribution and time delays is 
illustrated in Fig.~\ref{fig:meantheta_vs_jet_opening_angle} for different source distances. 
Photons emitted with a large angle $\theta_e$ with respect to the line of sight are typically 
observed with a large angle $\theta \approx \theta_e / \tau_{\gamma\gamma}$. As a result, the 
averaged detection angle in the 1 GeV - 100 TeV band is expected to scale as $\langle \theta 
\rangle\ \propto\ \theta_{\rm jet}$. This is clearly observed for distant sources with $z>0.4$ at 
small angle. The increase of the average angle with the jet opening angle is less pronounced for 
nearby sources. This results from a combination of the source intrinsic spectrum (different 
primary energies correspond to different annihilation distances) and the complex absorption 
feature for nearby sources (see Fig.~\ref{fig:spectrum_vs_redshift} and the associated 
discussion). At larger jet opening angle, the average detection angle is limited by the physical 
extension of the halo around the sources. For all opening angles, the average angle decreases 
with the source redshift mostly because of geometrical effect (halos look smaller when
observed from larger distances). 

The average time delay increases with the jet opening angle. Indeed with a large jet opening,
there is a larger halo effect. Then photons are more deflected and arrive at latter times. 
The redshift dependence is less pronounced than that of the average angle, as already mentioned.  

These results show that the effect of the source beaming on the cascade properties cannot
be modelled by simple one-generation models, and that it must be investigated numerically.

\section{Properties of the intergalactic medium}
\label{sec:intergalactic_medium}

In section \ref{sec:source_profile} we explored the effect of the source parameters on the
development and observability of a cosmological cascade. We now illustrate the impact 
of the intergalactic medium, in particular the effects of:
\begin{itemize}
   \item the extragalactic background light model
   \item the extragalactic magnetic field (amplitude $B$ and coherence length $\lambda_B$)
\end{itemize}
The same fiducial simulation is used, as in section \ref{sec:source_profile}.

\subsection{Extragalactic background light}
\label{sec:EBL}

\begin{figure} \centering
   \includegraphics[width=\columnwidth]{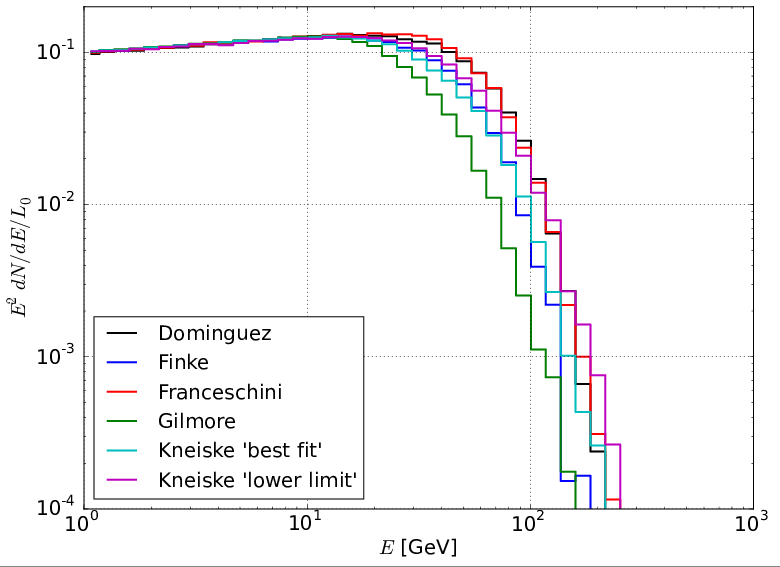}
   \caption{Full-sky spectrum for different EBL models. The source is
   located at redshift $z=2$.}
   \label{fig:spectrum_vs_EBL}
\end{figure}

The extragalactic light (EBL) affects the absorption of gamma-rays. Fig.~\ref{fig:spectrum_vs_EBL} 
shows the spectrum computed for an isotropic source with a spectral index of $\Gamma=1.2$ at a 
redshift $z=2$ using 6 different EBL models \citep{franceschini_extragalactic_2008,
dominguez_extragalactic_2011, finke_modeling_2010, kneiske_lower-limit_2010, 
gilmore_semi-analytic_2012}. The different EBL models predict different cut-off energies. 
This dependence on EBL models is similar to the analytical expectation from Fig. 
\ref{fig:cutoff_energy}. Below the cutoff energy, the cascade spectrum is however quite 
universal (in shape and intensity). Indeed the intrinsic spectrum used here is hard so that 
most of the absorbed energy corresponds to photons with energy close to the maximal energy 
$E_{\rm max}$ of the intrinsic spectrum, independently of the absorption energy $E_{\rm abs}$. 

\begin{figure} \centering
   \includegraphics[width=\columnwidth]{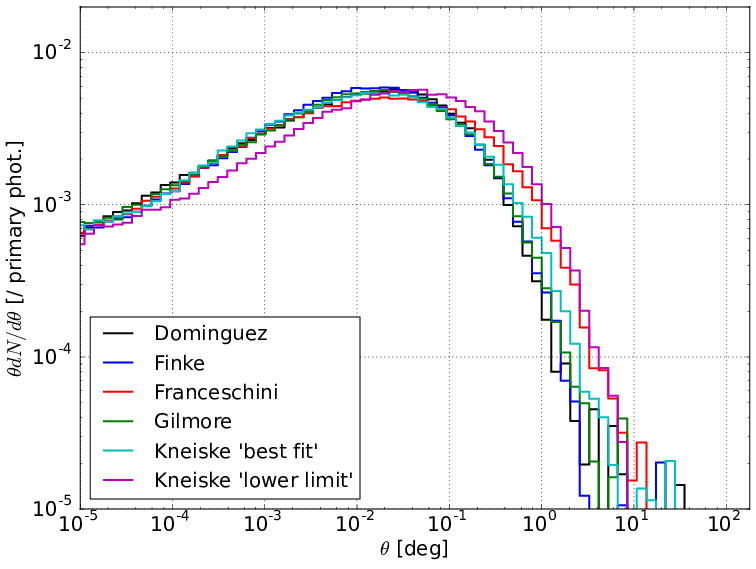}
   \caption{Detection angle distribution of $E>1$ GeV photons emitted at $z=2$ and for different 
   EBL models.}
   \label{fig:dNdtheta_vs_EBL}
\end{figure}

Fig.~\ref{fig:dNdtheta_vs_EBL} shows the distribution of detection angles for photons with
energies $E>1$ GeV and for the 6 different EBL models. 
Although the general shape of the distribution is similar from one model to the other, the typical 
angular scales involved can vary by a factor from 1 to 6.

\subsection{Extragalactic magnetic field}

The amplitude and coherence length of EGMF have no effect on the full integrated spectrum 
over an infinite time. It can however have an effect if the spectrum is integrated only over a
finite observational time or if a limited aperture is used. These questions have already been 
largely studied in the literature \citep[e.g.][]{taylor_extragalactic_2011,vovk_fermi/lat_2012,
arlen_intergalactic_2014}. We do not reproduce these studies here.

\begin{figure}
   \includegraphics[width=\columnwidth]{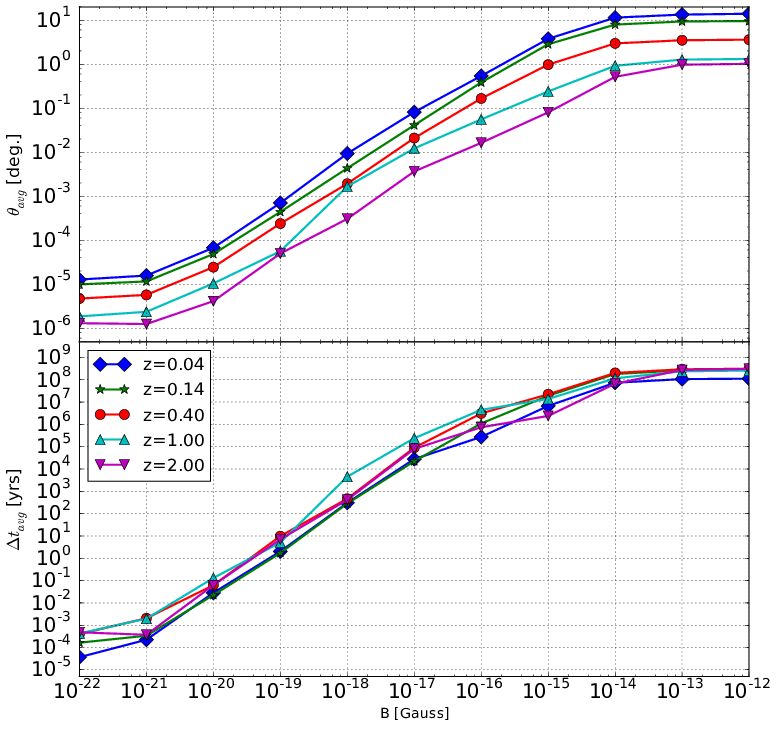}
   \caption{Average detection angle (top) and average time delay (bottom) of photons with energy 
      $E > 1$ GeV as a function of the EGMF strength for different source redshifts. The coherence 
      length is $\lambda_B=1$ Mpc.}
   \label{fig:mean_Dt_theta_vs_EGMF}
\end{figure}

Fig.~\ref{fig:mean_Dt_theta_vs_EGMF} shows the average time delay and arrival angle of 
photons with energy $E > 1$ GeV versus the amplitude of the EGMF for a coherence length 
$\lambda_B= 1$ Mpc. 
For strong magnetic fields ($B>10^{-14}$ G), leptons are trapped near their production site by the 
magnetic field and produce an isotropic source of typical size $\lambda_{\gamma\gamma}$. This 
translates into a typical average angle (and time delay), that is independent of the field 
intensity \citep{aharonian_very_1994}. For weak magnetic fields ($B <10^{-14}$G), there is no 
isotropisation of the emission and the average angle increases with the magnetic field intensity: 
the stronger the field, the larger the deflection and the larger the detection angle 
\citep{elyiv_gamma-ray_2009}. For very weak magnetic fields ($B <10^{-21}$G, the arrival angle and 
time delay saturate at minimal values corresponding to the intrinsic extension of the cascade, 
resulting from the small misalignment of the product particles with respect to their parent 
particles during pair production and Compton interactions. This intrinsic extension is independent 
on the magnetic field strength \citep[Eq. 41]{neronov_sensitivity_2009}. Also, nearby sources 
naturally appear as more extended than distant sources. Nonetheless, as for the effect of the 
jet opening angle, a combination of the intrinsic spectrum of the source and the complex 
absorption profile at low redshift produces an increase of the average angle that can be slower 
than the linear expectation of Eq. \ref{eqn:theta} and that depends on the source redshift.

\begin{figure}
   \includegraphics[width=\columnwidth]{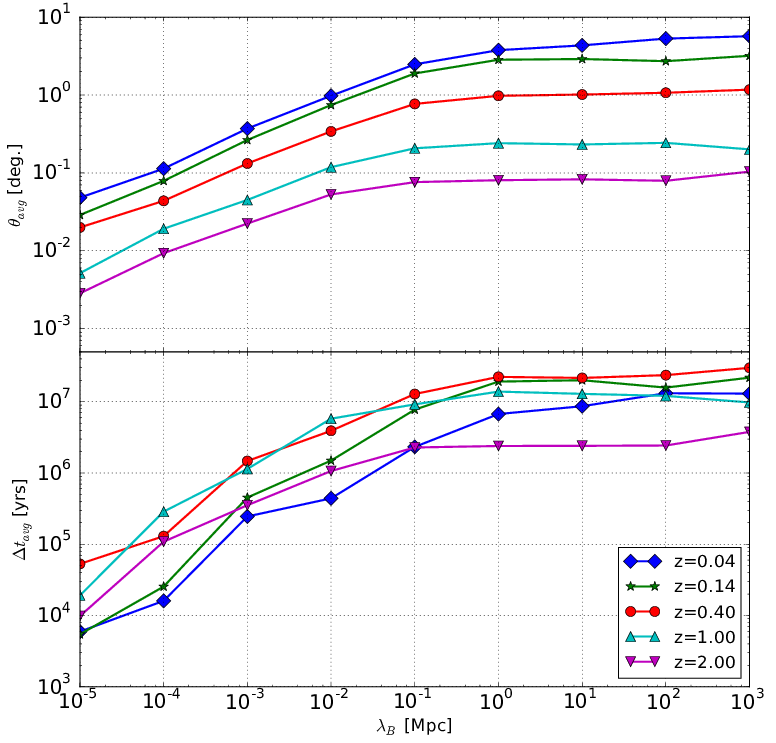}
   \caption{Average detection angle (top) and average time delay (bottom) of photons with energy 
      $E > 1$ GeV as a function of the EGMF coherence length $\lambda_B$ for different source 
      redshifts. The field strength is $B=10^{-15}$ G.}
   \label{fig:mean_Dt_theta_vs_LB}
\end{figure}

The properties of the cascade emission also depend on the coherence length. The general shape of 
the angular and time delay distributions (not shown here) are quite insensitive to that length 
scale, but the typical angular (and time) scales depend on it. The average angle and time delays 
of photons with energy $E > 1$ GeV are shown as a function of the coherence length in 
Fig.~\ref{fig:mean_Dt_theta_vs_LB}.
At large coherence length ($\lambda_B > 1$ Mpc), the leptons only visit a single uniform magnetic 
cell and the deflection is governed by the orientation of the field in that cell. This produces a 
halo size that is quite independent of the coherence length. At small coherence length ($\lambda_B 
< 10$ kpc), the leptons travel through many magnetic cells and have a random walk. The deflection 
angle then increases as $\lambda_B^{1/2}$, and so does the detection angle. Although the angular 
scales cannot be derived from the simple one-generation model, the expected behaviour  
$\theta_{avg}\propto\lambda_B^{1/2}$ is well recovered by the full simulations. In the simple 
model presented in section \ref{sec:observables}, the transition between the two regimes occurs 
when the lepton cooling distance associated to the observation energy $E$ equals the magnetic 
coherence length. We consider only photons above 1 GeV and at $z=0.14$, the absorption energy is 
about 1 TeV. The corresponding cooling distance of the parent leptons is then in the range 
$10<D_{ic}<500$ kpc (according to eq. \ref{eqn:Egamma} and \ref{eqn:Dic}). The transition is then 
expected to occur in the same range $10<\lambda_B<500$ kpc as observed in Fig 
\ref{fig:mean_Dt_theta_vs_LB}. Whereas the absorption energy depends only moderately on the 
source distance, the cooling distance scales as $D_{ic} \propto (1+z)^{-4}$, so that the 
transition occurs at much smaller coherence length for high-z sources, as observed in Fig. 
\ref{fig:mean_Dt_theta_vs_LB}.

\section{Conclusion}
\label{sec:conclusion}

In this article, we have presented a new, publicly available, Monte Carlo code to model the 
emission of electromagnetic cascades. This code makes very few approximations: it uses the exact 
interaction cross sections, both the EBL and CMB photons are targets for the lepton and photon 
interaction, it computes the exact 3D trajectory of leptons in cubic cells immersed in an uniform 
magnetic field and it takes into account the cosmological expansion in the evolution of the target 
properties, the particle energy and the particle trajectories. It can model the emission of 
cascades initiated by sources at any distance (as long as EBL models are accessible), with any 
intrinsic spectrum and any axisymmetric emission. The code was validated by comparison to 
published results and analytical estimates.

With this code, we have studied the role of the different physical parameters (source
spectrum, redshift, anisotropic emission, EBL spectrum and EGMF) involved in the cascade 
properties. This study also emphasises the limitations of the analytical estimates often 
used in the interpretation of high-energy observations. In particular, high-generation photons 
quickly dominate the cascade properties as soon as the intrinsic spectrum extends to energies 
significantly larger than the source absorption energy (typically a few TeV). 
Most studies use one-generation models to interpret the results of search for pair halos. 
We have shown that the angular distribution at potentially observable scales ($\theta>0.1^\circ$) 
can be fully dominated by second-generation photons if the intrinsic spectrum of the source is 
hard and extends to high energy. In that case, analytic expressions and their interpretations 
can be misleading.

The dependence of cosmological cascades on the characteristics of the sources (variability, 
spectrum, anisotropy) as well as those of the intergalactic medium (EBL, EGMF) is subtle and 
complex. Moreover, the observable properties of cascades depend also crucially on the 
way observations are conducted (aperture angle of the instrument, energy bands, time of the 
observation and time of exposure\dots). A detailed numerical modelling of the cascades is a 
prerequisite to disentangling all these effects. The code can be used: to interpret data, to
design observational strategies for present and future high-energy telescopes and to tighten 
the constraints on the EGMF. 

\section*{Acknowledgments}
This work has been carried out thanks to the support of the OCEVU Labex (ANR-11-LABX-0060) and the
A*MIDEX project (ANR-11-IDEX-0001-02) funded by the "Investissements d'Avenir" French government
program managed by the ANR. The authors are indebted to Guillaume Dubus for many useful comments 
and suggestions.

\bibliographystyle{mnras}
\bibliography{references}

\appendix
\section{Monte Carlo code}
\label{appendix:code}

The code was designed to be accurate and general, without any approximation, fast, and easy to use. 
It is written in \verb|Fortran 95| and parallelised with \verb|OpenMP|. The source code is 
available at the following address: \url{https://gitlab.com/tfitoussi/cascade-simulation}
with some additional informations. Simulations return series of events (detected photons)
which are then post-processed to generate observables such as spectra, angular distributions, 
images or timing delays. Post-processing scripts in \verb|python 2.7| are accessible at 
\url{https://gitlab.com/tfitoussi/simulation-analysis}. These scripts were developed for our own 
needs and should be considered as examples and templates. More details are given on the mentioned 
Git repositories.  

\begin{figure} \centering
   \includegraphics[width=\columnwidth]{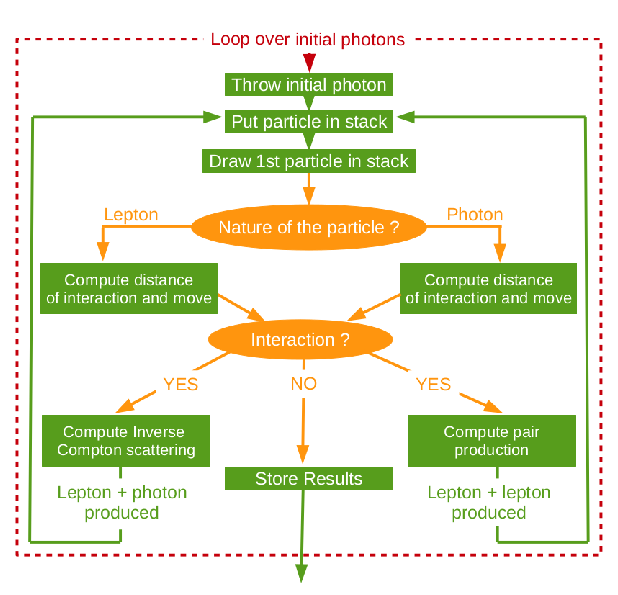}
   \caption{Scheme of the simulation process.}
   \label{simulation_process}
   \label{fig:algo}
\end{figure}

The general architecture of the code is presented in Fig.~\ref{fig:algo} (see also 
section~\ref{sec:code}), additional details are given below on the particular way some selected 
physical or numerical issues are dealt with.

\subsection{Particles propagation}
\label{appendix:travel}

In this section, we describe how the propagation of leptons is computed.

The Friedmann - Lemaitre - Robertson - Walker metric for a spatially flat and isotropic universe 
is:
\begin{equation} \label{FLRW_metric}
   ds^2 = c^2 dt^2 - a^2(t) d\chi^2,
\end{equation}
where $t$ is the cosmic time, $a(t)$ is the scale factor and $\chi$ the comoving distance. 
The code uses a flat $\Lambda$-CDM concordance model ($\Omega_k=0$). The results presented in 
this paper have been obtained with $\Omega_M=0.3$, $\Omega_\Lambda=0.7$ and the Hubble constant 
$H_0=67.8$ km s$^{-1}$ Mpc$^{-1}$. The evolution of the Universe is described by the
Friedmann - Lemaitre equation:
\begin{equation}
   \label{FL_equation}
   \frac{H^2}{H_0^2} = \left( \frac{\dot{a}}{a} \right)^2 = \Omega_M (1+z)^3 + \Omega_k (1+z)^2 + 
      \Omega_\Lambda 
\end{equation}
where $z=1/a-1$ is the cosmological redshift. Then, the evolution of redshift with cosmic time is 
governed by:
\begin{equation} \label{dt}
   dt = \frac{da}{\dot{a}} =  \frac{-dz}{(1+z) H(z)}.
\end{equation}
Photons follow geodesics $ds^2=0$. The comoving distance travelled by photons between two 
interactions is therefore:
\begin{equation}  \label{eq:dchi}
   \Delta \chi =  c \int_{t_1}^{t_2} \frac{dt}{a} = c \int_{z_2}^{z_1} \frac{dz}{H(z)} .
\end{equation}

The lepton propagation is affected by the extragalactic magnetic field (EGMF) and their equation 
of motion is:
\begin{equation}
   \label{motion_equation}
   \frac{d^2 x^\mu}{d\tau^2} + \Gamma^\mu_{\ \alpha \beta}
   \frac{dx^\alpha}{d\tau} \frac{dx^\beta}{d\tau} =
   \frac{ f^\mu}{m_e},
\end{equation}
where $m_e$ is the lepton mass, $x^\mu$ is the position quadri-vector, 
$\tau$ is the particle proper time, the $\Gamma^\mu_{\ \alpha \beta}$ are the Christofell symbols, 
and $f^\mu$ is the Lorentz force computed from the electro-magnetic tensor. We assume a passive 
field diluted as $B(z)=B_0/a^2$ where $B_0$ is its value at $z=0$. In terms of comoving 
coordinates $\vec{\chi}$ and cosmic time $t$, the equations of motion read:
\begin{align}
 \frac{d^2t}{d\tau^2} +\frac{a}{c^2} \frac{da}{dt} \left(\frac{d\chi}{d\tau}\right)^2
      & = 0 ,\\
 \frac{d^2\vec{\chi}}{d\tau^2} +\frac{2}{a} \frac{da}{dt} \frac{dt}{d\tau} 
            \frac{d\vec{\chi}}{d\tau}
      & = \frac{e_c}{a^2 m_ec} \frac{d\vec{\chi}}{d\tau} \times \vec{B}_0
\end{align}
where $e_c$ is the lepton charge and $c$ the speed of light. The cosmic time is linked to the 
particle proper time as: $d\tau=dt/\gamma$ and the comoving velocity is related to the proper 
velocity $v$ as: $d\chi/dt = v/ac$. Defining $p=(\gamma^2-1)^{1/2}$ the lepton dimensionless 
momentum, the first equation reduces to $\dot{p}/p = -\dot{a}/a$, the solution of which describes 
how the particle energy evolves with cosmic time:
\begin{equation}
   p \propto 1/a = (1+z) .
\end{equation}
Note that highly relativistic particles behave very much like photons since $p\approx \gamma$. 
The second equation is simpler when written in terms of a {\it generalised} conformal time $\eta$ 
defined by $d\eta = dt/ (\gamma a^2)$:
\begin{equation}
   \frac{d^2\vec{\chi}}{d\eta^2} + \vec{\omega}_{B,0} \times \frac{d\vec{\chi}}{d\eta} = \vec{0},
\end{equation}
where $\vec{\omega}_{B,0} = e_c \vec{B}_0/(m_ec)$ is the cyclotron pulsation at $z=0$
(i.e. a constant). This equation is the equation of motion of a charged particle in a constant 
magnetic field. The solution is the classical helicoidal motion. The period of generalised conformal time 
between two interactions occurring at cosmic times $t_1$ and $t_2$ is computed numerically as:
\begin{equation}
   \Delta \eta = \int_{t_1}^{t_2} \frac{dt}{\gamma a^2} 
               = \int_{z_2}^{z_1} \frac{(1+z)}{\gamma(z) H(z)} dz \label{eq:deta}.
\end{equation}
And the associated rotation angle is:
\begin{equation}
\theta = -\omega_{B,0} \Delta \eta .
\end{equation}
The trajectory is then easily described by Rodrigues' rotation formulae. Let $\vec{b}=\vec{B}/B$ 
the direction of the magnetic field and $\vec{u}_1=\vec{v}_1/v_1$ the velocity direction at time 
$t_1$. Defining two orthogonal vectors in the perpendicular plane $\vec{e_1} = 
\vec{b}\times\vec{u}_1$ and $\vec{e_2}=\vec{b}\times \vec{e_1}$, the change in the velocity 
direction and in comoving position read respectively:
\begin{align}
   \Delta \vec{u} &= \sin\theta \vec{e_1} + 2\sin^2\left(\theta/2\right) \vec{e_2} \\
   \frac{\Delta \vec{\chi}}{\chi_s} &= \vec{u}_1 + \sin\left(\theta/2\right){\rm sinc}
   \left(\theta/2\right) \vec{e_1} + \left(1-{\rm sinc }~\theta\right)
   \vec{e_2}
\end{align}
where $\chi_s=c  \Delta \eta p/(1+z) $ is the curvilinear, comoving distance travelled by the 
lepton, and ${\rm sinc}(x)=\sin{(x)}/x$.

Integrals \ref{eq:dchi} and \ref{eq:deta} are computed numerically using a Gauss-Legendre 
quadrature with 30 quadrature points allowing to reach machine precision for floats in double 
precision.

\subsection{Accuracy}

Monte Carlo simulations of electro-magnetic cascades might suffer from numerical accuracy issues 
when very small angles and time delays are computed. The code is run with double precision floats, 
with relative precision of $\sim 10^{-15}$. This provides sufficient precision to deal with 
realistic cascades. 

The particle directions and positions are described numerically by 3D vectors. Angles are
typically computed from the arccosine of the dot product of two such vectors. The relative 
precision on angles is then about $\sim10^{-7}$ rad, which is far below any instrumental PSF.  

Time delays are computed by comparing the arrival time of cascade photons to the arrival time of 
primary photons. The particle motion is computed by integrating the trajectory, including the 
cosmological expansion (see Sec. \ref{appendix:travel}). This implies numerical integration and 
iterative methods. As explained above, numerical integrations use enough quadrature points to 
reach machine precision as long as $z<6$. In addition, the computation of the interaction distance 
is performed using the Newton-Raphson method with a convergence criterium small enough to also 
reach machine precision. 
For our cosmological model, the age of the universe is $4.3\times10^{17}$ s. The machine precision 
of $10^{-15}$ then allows to access time delays shorter than one hour, which is also enough for 
the typical integration time of realistic observations with {\it Fermi} and Cherenkov telescopes.

\subsection{Acceleration methods}

Although Monte Carlo simulations allow for an exact description of each particle motion and 
interaction, more and more particles must be tracked by the code as the cascade develops. Here we 
describe some aspects of numerical acceleration methods that allow for enough cascade statistics 
and fast computation time, with negligible effect on the cascade physics. We use three methods to 
reduce the computing time.


\begin{enumerate}
\item As leptons cool down, they end up producing Compton photons at low energy, when they 
   are not expected to be observed because of strong sky background and low instrument 
   sensitivity. Below some threshold energy, leptons are thus removed from the stack and are not 
   tracked further by the code. Identically, photons that are not detected are discarded below a 
   corresponding threshold energy. In the results presented here, we used a lepton threshold 
   energy of $E_{\rm th}^e = 5.56$ GeV, which corresponds to describing the cascade emission only 
   above 100 MeV.

\item As the cascade develops, many low-energy particles are produced, providing a much better 
   statistics at low energy. In order to increase the relative statistics at high energy, a 
   weighted sampling is performed at each interaction, based on the energy of the involved particles. 
   Namely, only a fraction $f_{s} = (E_{\rm out}/E_{\rm in})^{\alpha_s}$ of each outgoing
   particle is kept, where $E_{\rm out}$ and $E_{\rm in}$ are the energy of the child and 
   parent particles respectively, and where $\alpha_s$ is an efficiency parameter (typically 
   $\alpha_s=0.6$ for the results shown here). In order to keep track of the physical number of 
   particles in the simulation, each particle is given a weight. And at each interaction, the 
   weight of the particles that are not discarded by this sampling method is increased by a 
   factor $f_s$. This procedure is applied both to leptons and photons.

\item As the typical Compton interaction length ($\lesssim 1$ kpc) is much shorter than the photon 
   annihilation distance ($\gtrsim 1$ Mpc), most of the time is spent computing the motion and 
   Compton scatterings of leptons. At low energy, leptons up-scatter a huge number of target 
   photons to gamma-ray energies. This large number of interactions at low energy allows to 
   perform a sampling of the Compton interactions. Namely, target photons are gathered in 
   macro-photons consisting of $N$ physical photons (not necessary an integer number). Leptons 
   are then considered to interact with this much sparser population of targets, which spares 
   a significant fraction of computation time. The interaction distance, lepton energy 
   loss and weight of the produced gamma-rays are increased accordingly with factor $N$ 
   while the deflection angle is increased by a factor $N^{1/2}$. The size of the macro-photons 
   is chosen such that on average a constant fraction $\eta$ of the lepton energy is typically 
   lost during each interaction with a macro-photon: $N=\eta E_k / \langle \Delta E
   \rangle$, where $E_k=(\gamma-1)m_ec^2$ is the lepton kinetic energy, $\langle \Delta E
   \rangle =(4/3)(\gamma^2-1) E_{cmb}$ is the average lepton energy loss in the Thomson 
   regime at each scattering event on CMB photons, and $E_{cmb}=2.7 T_{cmb}$ is the average energy 
   of CMB photons. We used $\eta=0.5\%$ for the results presented here. Such an interaction 
   sampling is performed only when $N>1$, that is for leptons with kinetic energy lower than 
   $E_k/(m_ec^2) = (3/4) \eta m_ec^2/E_{cmb}-2$, that is about $E_k<1$ TeV at $z=0$. In practice, 
   the numerical computation of the lepton energy after the up-scattering of a macro-photon can 
   lead to negative energies for several reasons (energy 
   dispersion around the average value, interactions with EBL photons, interactions in the 
   Klein-Nishina regime, and overestimation of the energy variation by assuming a constant lepton 
   energy). However, such events remain very rare given the small fraction $\eta$ chosen here, and 
   are simply discarded. As for particle sampling, the Compton sampling increases the relative 
   statistics at high energy, where only few photons are detected.
\end{enumerate}

\subsection{Modelling isotropic and anisotropic emission}
\label{appendix:anisotropy}

\subsubsection{Code outputs}
Regardless of the geometrical properties of the source emission, the code is always run with 
primary photons emitted in a single direction $\vec{e}_z$. Secondary photons are detected when 
they cross the sphere of comoving radius equal to the comoving distance to the source. There, 
the following properties of each detected photon $i$ are recorded: 
\begin{itemize}
   \item[-] Weight $w_i$
   \item[-] Energy $E_i$
   \item[-] Time delay $\Delta t_i$
   \item[-] Angular position on the sphere $\vec{n}_{p,i}$. Positions are measured in a
      spherical system of coordinates ($\theta_{p,i}^{\rm code},\phi_{p,i}^{\rm code}$)
      with main axis $\vec{e}_z$ (as we will see, we consider only axisymmetric cascades, 
      so that the orientation of this base around $\vec{e}_z$ is irrelevant). 
   \item[-] Angular direction $\vec{n}_{d,i}$. Directions are measured from an observer
      point of view in a spherical system of coordinates ($\theta_{d,i}^{\rm
      code},\phi_{d,i}^{\rm code}$) with main axis the position vector $\vec{n}_p$ and
      orientation defined by the direction of the primary photons $\vec{e}_z$.
\end{itemize}

\subsubsection{Physical observables}
From these outputs, the case of an {\it isotropic source} is simply modelled by means of
rotations. Namely, the integration over all emission directions $\vec{n}_e$ with respect to 
the line of sight is equivalent to the integration over all detection positions in the code 
outputs. Hence photons detected at any position $\theta_p^{\rm code}$ are kept with their other 
properties unchanged (weight, energy, time, direction) and any distribution (e.g. energy spectrum, 
angular distribution of detected photon etc) can be derived. 

The case of {\it anisotropic emission} is more complex to handle. Here is a short description of
the method used to model any axisymmetric emission in a post-processing stage. The
derivation is based on two main assumptions:
\begin{itemize}
   \item[-] First, it is assumed that the source emission is axisymmetric around a jet
      axis that makes an angle $\theta_{\rm obs}$ with respect to the line of sight i
      (see Fig. \ref{fig:triangle}). The
      source emission is described by its angular distribution $dN_e/d\Omega_e$, where
      $d\Omega_e = -\sin\theta_e d\theta_e d\phi_e$ is the elementary solid angle and the
      spherical coordinates $(\theta_e,\phi_e$) are defined with respect to the jet axis
      $\vec{n}_{\rm jet}$. The emission distribution is typically characterised by its
      half opening angle $\theta_{\rm jet}$. For instance, 
      the disk profile used in Fig. \ref{fig:image_tobs} is a uniform
      angular distribution up to the jet half opening angle: $dN_e/d\Omega_e  = 1/(2\pi
      (1-\cos\theta_{\rm jet}))$ if $\theta_e < \theta_{\rm jet}$ and 0 otherwise.
   \item[-] Second, it is assumed that the cascade initiated by a single primary photon is
      axisymmetric around the direction of this primary. This approximation is accurate
      for small magnetic coherence lengths for which many different magnetic cells with
      random orientations are crossed by the particles. It might become less accurate when
      the magnetic field is coherent over very large distances. However, our numerical
      simulations show that cascades initiated by unidirectional primary photons of 
      $E_{\gamma,0}<100$ TeV, emitted at $z=0.13$ remain highly axisymmetric up to 
      $\lambda_B=10$ Mpc. Although images of misaligned jets might be inaccurate above 
      this scale, such limitation vanishes for aligned jets and isotropic emission. 
\end{itemize}

The code provides the distribution of photons in a local system of coordinated based on the direction of each primary photons. To recover the total source geometry, we must transform these results to a distribution of photons in a global system of coordinates. Energy and time delay does not depend on the choice of coordinates are kept unchanged. The position of detected photons does depend on the choice of coordinates. However a given observer corresponds to one specific position so that the detection positions is a fix parameter. Hence only the direction of detected photons must be transformed carefully.

In the global frame, the direction of secondaries is best measured in a system of spherical coordinates $(\theta_d,\phi_d)$ with main axis the position vector $\vec{n}_p$ (i.e. the line of sight) and oriented with respect to the jet axis $\vec{n}_{\rm jet}$. The direction of this base is the same as the one used to measure directions in the code (hence $\theta_{d}=\theta_{d}^{\rm code}$). However, the orientation is now defined independently of the primary direction. Images are obtained by plotting the photon density in the sky
plane ($\theta_d \sin{\phi_d},\theta_d \cos{\phi_d})$. In the main body of the paper, the 
subscript $d$ has been dropped with no ambiguity.

Because of the axisymmetric assumptions, each detected photon actually produces a (non-uniform) ring in
the sky plane of the observer. Hence, the numerical result is a continuous function of $\phi_d$ and a discrete
function of $\theta_d$. For an observer misaligned by an angle $\theta_{\rm obs}$ from the
jet axis, the emission from photon $i$ observed at angle $\phi_d$ corresponds to primary
light emitted with an angle $\theta_{e,i}(\phi_d)$ away from the jet axis, and this angle
is defined as:
\begin{equation}
   \cos \theta_{e,i} = \cos \theta_{\rm obs} \cos \theta_{p,i}^{\rm code} + \sin
   \theta_{\rm obs} \sin \theta_{p,i}^{\rm code} \sin(\phi_{d}-\phi_{d,i}^{\rm code}).
\end{equation}
Since for an anisotropic source, the emission intensity depends on that angle, the weight
of each ring $i$ depends on the azimuthal angle $\phi_d$:
\begin{equation}
   W_{i}(\phi_d) =  \frac{w_i}{2\pi} \left. \frac{dN_e}{d\Omega_e}\right|_{\theta_{e,i}}.
\end{equation}
This procedure provides the number of photons detected per unit direction-solid angle and
per unit position-solid angle: $d^2N/(d\Omega_pd\Omega_d)$ at position
$\theta_p=\theta_{\rm obs}$ and in direction $(\theta_d,\phi_d)$. Dividing this further by
the luminosity distance $D_L$ of the source gives the surface brightness, that is the
number of detected photons per unit detector surface and per unit direction solid angle:
$dN/(dSd\Omega_d)= dN/(d\Omega_p\Omega_d)/D_L^2$.

\bsp	
\label{lastpage}

\end{document}